\documentclass[letterpaper,english,aps,prl,reprint,superscriptaddress]{revtex4-1}
\usepackage[latin9]{inputenc}
\usepackage{color}
\usepackage{amsmath}
\usepackage{amssymb}
\usepackage{graphicx}

\makeatletter

\pdfpageheight\paperheight
\pdfpagewidth\paperwidth

\usepackage{hyperref}
\hypersetup{
    colorlinks = true,
	allcolors = blue	
}
\usepackage{times}

\makeatother

\usepackage{babel}
\begin{document}
\preprint{APS/123-QED}
\title{
Two-particle Berry phase mechanism for Dirac and Majorana Kramers pairs of corner modes}
\author{Yi Tan}
\thanks{YT and ZHH contribute equally to this work.}
\affiliation{International Center for Quantum Materials and School of Physics, Peking University,
Beijing 100871, China}
\affiliation{Collaborative Innovation Center of Quantum Matter, Beijing 100871, China}
\author{Zhi-Hao Huang}
\thanks{YT and ZHH contribute equally to this work.}
\affiliation{International Center for Quantum Materials and School of Physics, Peking University,
Beijing 100871, China}
\affiliation{Collaborative Innovation Center of Quantum Matter, Beijing 100871, China}
\author{Xiong-Jun Liu}
\thanks{Corresponding author: xiongjunliu@pku.edu.cn.}
\affiliation{International Center for Quantum Materials and School of Physics, Peking University,
Beijing 100871, China}
\affiliation{Collaborative Innovation Center of Quantum Matter, Beijing 100871, China}
\affiliation{CAS Center for Excellence in Topological Quantum Computation, University of Chinese
Academy of Sciences, Beijing 100190, China}
\affiliation{Institute for Quantum Science and Engineering and Department of Physics, Southern
University of Science and Technology, Shenzhen 518055, China}
\begin{abstract}
We uncover a novel two-particle Berry phase mechanism to realize exotic corner modes in second-order
topological insulators (TIs) and topological superconductors (TSCs) with time-reversal symmetry. We show that the nontrivial pseudospin textures of edge states in two different types of two-dimensional TIs give rise to the novel two-particle geometric phases in the particle-hole and particle-particle channels, respectively, for which the edge mass domain walls or intrinsic $\pi$-junctions emerge across corners when an external
magnetic field or $s$-wave superconductivity is considered, hosting Dirac or Majorana corner modes. Remarkably, with this mechanism we predict the Majorana Kramers pair of corner modes by directly coupling the edge of a type-II time-reversal invariant TI to a uniform $s$-wave SC, in sharp contrast to the previous proposals which rely on unconventional SC
pairing or complex setting for fine-tuned SC $\pi$-junction. We find Au/GaAs(111) to be a realistic material candidate for realizing such Majorana Kramers pair of corner modes.
\end{abstract}

\maketitle
$\textit{Introduction}$.\textemdash{} Topological phases have generated great interest
in the past decades \citep{HasanRMP2010,QiZhangRMP2011,BansilRMP2016}. Recently,
the higher-order topological insulators (TIs) and topological superconductors (TSCs)
were proposed \citep{BenalcazarScience2017,BenalcazarPRB2017,LangbehnPRL2017,ZhidaSongPRL2017,SchindlerScience2018,GeierPRB2018,MiertPRB2018}.
An nth-order TI (TSC) in $d$ dimension ($d$D) features topologically protected
gapless states on its ($d-n$)D boundary, but is gapped elsewhere. For example, a
2D second-order topological insulator (SOTI) hosts in-gap modes at the corners, while
its both 2D bulk and 1D edges are gapped \citep{SAYangPRL2019,SAYangPRL2020}. The
physics of corner modes is readily understood from the perspective of Dirac equations:
The original helical edge states of a 2D time-reversal (TR) invariant TI are gapped
by a sign-changing Dirac mass, leading to the corner modes at the mass domain walls
\citep{JackiwRebbiPRD1976,CalugaruPRB2019,SchindlerJAP2020}. The same picture applies
to the case of 2D second-order topological superconductors (SOTSCs).

While the interests grow rapidly on the higher-order topological phases, the experimental
realization of such phases in solid materials is rare, albeit various interesting
theoretical schemes have been proposed \citep{SchindlerScience2018,EzawaPRB2018,LiuFengPRL(2019),SAYangPRL2019,ZFWangNanoLett2019,BJYangNPJ2020,SAYang-arXiv,NiuQianPRL2020,SAYangPRL2020,LiuFengPRResearch2021,ZFWangPRL2021}.
In comparison with higher-order topological insulators, the superconducting counterparts
are usually even more demanding in the realization \citep{LiuPRB2018,ShapourianPRB2018,YanPRL2019,ZhuPRL2019,WuPRB2019,ZhangPRL2019,YangPRB2019,PanPRL2019,RoyPRB2020}. In particular, for the second-order time-reversal (TR) invariant TSCs which host Majorana Kramers pairs (MKPs) of corner modes, the existing proposals rely on either the coupling of TI helical edge states to unconventional superconductivity~\citep{WangPRL2018,YanPRL2018}
or complex settings for fine-tuned SC $\pi$-junctions \citep{VolpezPRL2019,ZengPRL2019,Laubscher2020}.
The MKPs obey the symmetry-protected non-Abelian statistics, manifesting a new type of non-Abelian anyons~\citep{LiuPRX2014,localmixing1,localmixing2,GaoPRB2016,TRTSCReview2019,Knapp2020,Hong-arXiv}. Thus to uncover experimental schemes, albeit being currently very scarce~\cite{ZWang2021PRX}, to realize the MKPs is highly significant and of broad interests.

In this Letter, we propose a fundamental mechanism by introducing
a novel concept of two-particle Berry phase to characterize and realize topological
corner modes in second-order topological phases. We show that two different scenarios
of Berry phases arise from the pseudospin textures of edge states \citep{LiuPRL2014}
in two different types of 2D TIs, leading to the sign changes in the gap
opened at edge and across corners when external magnetic field or superconductivity
is applied. In the SOTI phase, the mass domain wall emerges and originates from the two-particle
Berry phase of particle-hole channel, leading to the transition from QSH to SOTI
phase by applying a uniform in-plane magnetic field. In the SOTSC phase, {\em intrinsic} $\pi$-junctions emerge at corners and are resulted from the two-particle Berry phases of particle-particle channel
(Cooper pair), giving rise to MKPs of corner modes with TR symmetry. Remarkably, with this mechanism we predict that the MKPs can be realized
by directly superposing a type-II TI on a uniform $s$-wave SC, in sharp contrast
to the previous proposals \citep{WangPRL2018,YanPRL2018,VolpezPRL2019,ZengPRL2019,Laubscher2020}.
This prediction updates a traditional viewpoint that, due to the TR symmetry, a
uniform $s$-wave proximate SC can induce a uniform pairing order without node in and thus fully gap out the helical edge states of TI, see e.g. refs~\citep{LFu2009,JasonRep2012}. 
Our finding shows that for a broad class of the quantum spin-Hall
insulators, 
the intrinsic $\pi$-junction emerges across corners when the edge is coupled to a uniform $s$-wave SC, leading
to the MKPs of corner modes. We show the QSH insulator Au/GaAs(111) thin film \citep{HilnerAPL2006,WangNatCommun2016} as
a 2D material candidate for the realization of the MKPs, and further provide
a generic guideline in searching for such SOTSC phases.

\begin{figure*}
\begin{centering}
\includegraphics{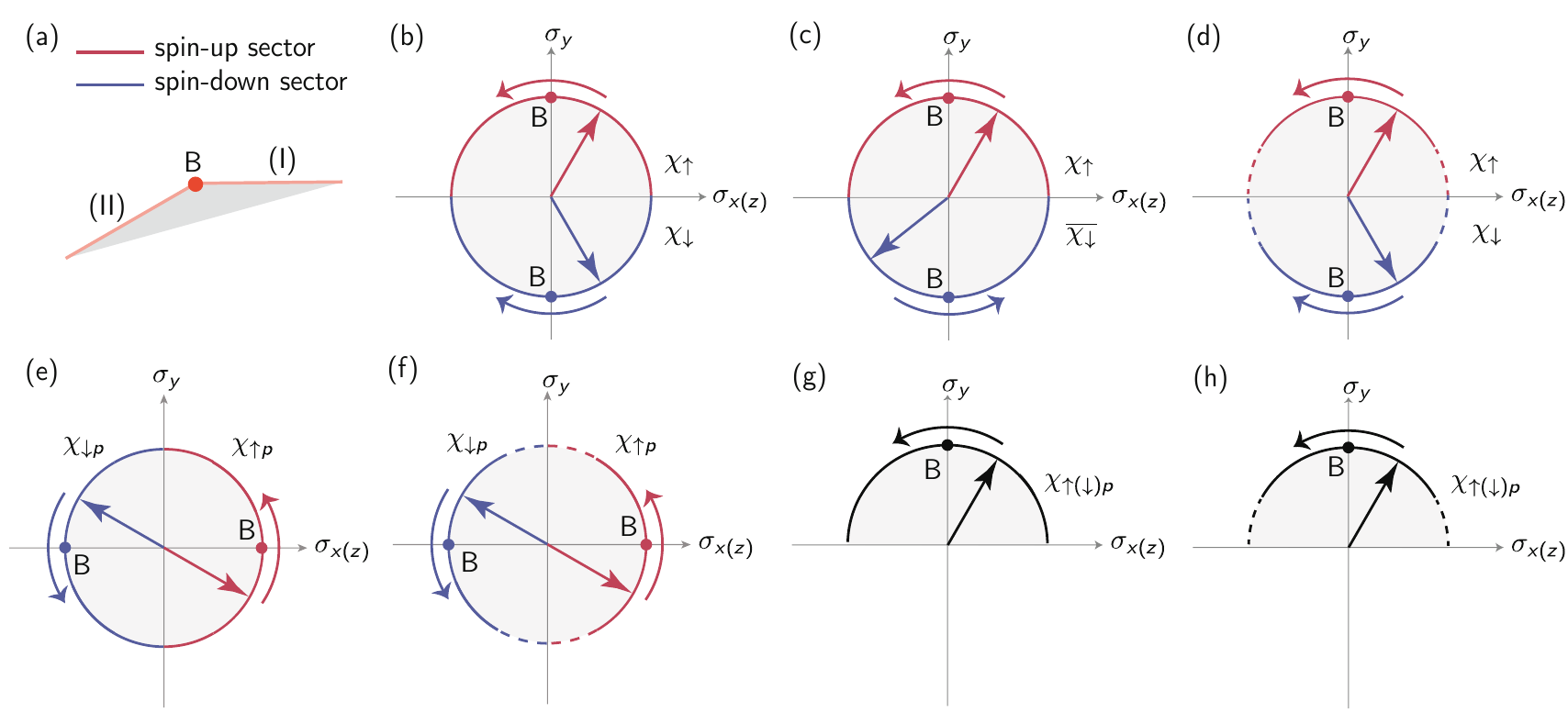}
\par\end{centering}
\caption{\label{fig:1}Boundary pseudospin textures of TIs crossing corner B. (a) Schematic
plot of the adjacent edges in real space. (b) Pseudospin trajectories for $|\chi_{\uparrow}\rangle$
(red solid lines) and $|\chi_{\downarrow}\rangle$ (blue solid lines) wind oppositely
in the intrinsic space. The straight red(blue) arrows denote the pseudospin vectors
for $|\chi_{\uparrow}\rangle$($|\chi_{\downarrow}\rangle$). (c) The joint trajectory
for $|\chi_{\uparrow}\rangle$ and $\overline{|\chi_{\downarrow}\rangle}$ can be
viewed as a closed loop in intrinsic space, which enables a well-defined two-state
Berry phase in PH channel $\gamma_{ph}$, leading to the Dirac corner modes. (d)
The conclusion in (c) also apply to the situation where pseudospin trajectories are
non-closed and can be connected by the continuation trajectories (dashed lines).
(e) and (g) The winding of pseudospin trajectories within both spin sectors are in
the same direction. The straight black arrows represent the coinciding of pseudospin
vectors for $|\chi_{\uparrow}\rangle$ and $|\chi_{\downarrow}\rangle$. And the
two-state Berry phase in PP channel $\gamma_{pp}$ is of importance since the joint
trajectories for $|\chi_{\uparrow p}\rangle$ and $|\chi_{\downarrow p}\rangle$
are closed, guaranteeing the appearance of the MKPs. This result can be extended
to non-closed cases (f) and (h).}
\end{figure*}

$\textit{The generic edge theory}$.\textemdash We start with a generic edge theory
of the geometric phase mechanism for corner modes. The theory is extended from a
previous result that the chiral edge state in a class of quantum anomalous Hall (QAH)
insulators exhibits topological (pseudo)spin texture in real space, with the pseudospin
polarization of each edge state winding one big circle along the closed 1D boundary in the presence
of chiral symmetry~\citep{LiuPRL2014,ZhangPRB2016}. Such pseudospin texture defines
a $\pi$ Berry phase on the boundary. For a TR invariant TI formed by two copies
of QAH insulators with opposite Chern numbers, its helical edge states form Kramers
pairs of the TRS $\mathcal{T}=is_{y}\mathcal{K}$, with $s_{y}$ being spin operator
and $\mathcal{K}$ the complex conjugate. The pseudospin polarizations of spin-up
($|\chi_{\uparrow}\rangle$) and spin-down ($|\chi_{\downarrow}\rangle$) edge states
are given by $\langle\boldsymbol{\sigma}_{\uparrow}\rangle=\langle\chi_{\uparrow}|s_{0}\boldsymbol{\sigma}|\chi_{\uparrow}\rangle$
and $\langle\boldsymbol{\sigma}_{\downarrow}\rangle=\langle\chi_{\downarrow}|s_{0}\boldsymbol{\sigma}|\chi_{\downarrow}\rangle$,
where the pseudospin operators $\boldsymbol{\sigma}$ denote the orbital (or sublattice)
degree of freedom. 
When traveling from one edge to a neighboring edge across a corner point B in the real space {[}Fig. \ref{fig:1}(a){]}, the pseudospin
direction ($\langle\boldsymbol{\sigma}_{\uparrow,\downarrow}\rangle$) of a single
state in general dose not trace a closed loop. 
However, a $\textit{two-state Berry phase}$ can be defined when the pseudospin polarizations
are inverted across the corner for both spin-up and spin-down sectors. Two basic
cases corresponding two fundamental types of TIs are of particular interests and studied below.

First, the pseudospin trajectories of the two spin sectors wind oppositely, e.g.
$\left(\langle\sigma_{\uparrow x}\rangle,\langle\sigma_{\uparrow y}\rangle\right)=\left(\langle\sigma_{\downarrow x}\rangle,-\langle\sigma_{\downarrow y}\rangle\right)$
{[}Fig. \ref{fig:1}(b){]}, so the joint pseudospin trajectory of the spin-up particle
and spin-down hole is closed in intrinsic space {[}Fig. \ref{fig:1}(c){]}. The case that TR symmetry reverses only real spin, but not pseudospin bases belongs to this scenario. We define
the two-state Berry phase in particle-hole (PH) channel
\begin{equation}
\gamma_{ph}=\int_{\theta^{(\mathrm{I})}}^{\theta^{(\mathrm{II})}}\mathrm{d}\theta\ \langle\chi_{\uparrow}(\theta)|\otimes\overline{\langle\chi_{\downarrow}(\theta)|}(-i\partial_{\theta})|\chi_{\uparrow}(\theta)\rangle\otimes\overline{|\chi_{\downarrow}(\theta)\rangle}.\label{eq:1}
\end{equation}
Here $\theta^{(\mathrm{I},\mathrm{II})}$ stands for the normal direction of edge
I and II, and the spin-down hole state $\overline{|\chi_{\downarrow}\rangle}$
is obtained by charge conjugation $\mathcal{K}$ on $|\chi_{\downarrow}\rangle$,
equivalent to flipping a ket into a bra state. Applying a uniform
in-plane Zeeman field ($\mathbf{M}\cdot\mathbf{s})\sigma_{0}$ to break the TRS and
introduce mass term $\mathcal{H}_{\mathrm{mass}}$ \citep{SM} to the edge Dirac
Hamiltonian $\mathcal{H}_{\mathrm{edge}}=vk_{\parallel}s_{z}$ \citep{SAYangPRL2020}. The matrix element $\langle\chi_{\uparrow}|(\mathbf{M}\cdot\mathbf{s})\sigma_{0}|\chi_{\downarrow}\rangle$
resembles a particle-hole excitation gap on the edge, and when turning
from edge I to edge II across the corner point $B$, the mass term is related by
\citep{SM}
\begin{equation}
\langle\chi_{\uparrow}^{\mathrm{(II)}}|(\mathbf{M}\cdot\mathbf{s})\sigma_{0}|\chi_{\downarrow}^{\mathrm{\mathrm{(II)}}}\rangle=e^{-i\gamma_{ph}}\langle\chi_{\uparrow}^{\mathrm{(I)}}|(\mathbf{M}\cdot\mathbf{s})\sigma_{0}|\chi_{\downarrow}^{\mathrm{(I)}}\rangle.\label{eq:2}
\end{equation}
Consequently, for $\gamma_{ph}=\pm\pi$ a mass domain wall is obtained across the
corner, giving zero-energy corner modes. This result can be extended to the more generic configuration
that the pseudospin polarization of each spin sector is not fully inverted across
the corner {[}Fig. \ref{fig:1}(d){]}. In this case we consider a continuation {[}shown
in dashed line of Fig. \ref{fig:1}(d){]} to connect them so that the two-state pseudospin
trajectory is still closed and a two-state $\pi$ Berry phase is resulted. As long
as no node of mass is obtained on the continuation trajectories, the mass domain
wall must be obtained across the corner. We note that the $\pi$ Berry phase and zero-energy mode for the insulating phase necessitate protection by a chiral-like symmetry~\citep{SM}, which anticommutes with all terms in Hamiltonian. When this symmetry is broken, the pseudospin texture
will not be fully in-plane and the Berry phase is no longer strictly quantized to
$\pi$ (even with closed pseudospin trajectories via continuation), hence not guaranteeing
the full sign reversion of edge mass term. In this case the corner mode has finite energy but may still be within the gap if the Berry phase is close to $\pi$-value.

The second even more nontrivial case that we unveil is the two-state Berry phase in the particle-particle
(PP) channel, which corresponds to realizing MKPs of corner modes in SOTSCs when the edge is proximately
coupled to a uniform $s$-wave SC. The case that TR symmetry reverses both real spin and pseudospin bases belongs to this scenario. Particularly, when the pseudospins of both spin
sectors wind in the same direction across the corner, e.g. $\left(\langle\sigma_{\uparrow x}\rangle,\langle\sigma_{\uparrow y}\rangle\right)=\pm\left(\langle\sigma_{\downarrow x}\rangle,\langle\sigma_{\downarrow y}\rangle\right)$
{[}Fig. \ref{fig:1}(e,g){]}, the two-state Berry phase in PH channel vanishes, but
is nontrivial in PP channel
\begin{align}
\gamma_{pp} & =\int_{\theta^{(\mathrm{I})}}^{\theta^{(\mathrm{II})}}\mathrm{d}\theta\ \langle\chi_{\uparrow p}(\theta)|\otimes\langle\chi_{\downarrow p}(\theta)|\nonumber \\
 & \hspace{7em}(-i\partial_{\theta})|\chi_{\uparrow p}(\theta)\rangle\otimes|\chi_{\downarrow p}(\theta)\rangle.\label{eq:3}
\end{align}
For the edge Hamiltonian $\mathcal{H}_{\mathrm{edge}}=ivk_{\parallel}\tau_{y}s_{z}$,
the SC pairing term induced in the edge states reads $\mathcal{H}_{\mathrm{mass}}=\langle\chi_{\uparrow p}|\Delta_{s}\tau_{y}s_{y}\sigma_{0}|\chi_{\downarrow h}\rangle\tau_{y}s_{y}$,
where $\tau_{y}$ represents the particle-hole degree of freedom, and subscript $p/h$
denotes the Nambu particle/hole sector. This matrix element gives the SC gap function
in the edge and obeys \citep{SM}
\begin{equation}
\langle\chi_{\uparrow p}^{\mathrm{(II)}}|\Delta_{s}\tau_{y}s_{y}\sigma_{0}|\chi_{\downarrow h}^{\mathrm{\mathrm{(II)}}}\rangle=e^{-i\gamma_{pp}}\langle\chi_{\uparrow p}^{\mathrm{(I)}}|\Delta_{s}\tau_{y}s_{y}\sigma_{0}|\chi_{\downarrow h}^{\mathrm{(I)}}\rangle.\label{eq:4}
\end{equation}
In this case, a mass domain wall (an {\em intrinsic} $\pi$ junction) is obtained when $\gamma_{pp}=\pm\pi$
and harbors at each corner a MKP, which obeys symmetry-protected
non-Abelian statistics~\citep{LiuPRX2014,localmixing1,localmixing2,GaoPRB2016,TRTSCReview2019,Knapp2020,Hong-arXiv}. With the particle-hole symmetry being always present in superconductors, the MKPs of zero corner modes are protected by time-reversal symmetry, without the necessity of chiral-like symmetry in insulators. We emphasize that
in this realization we directly superpose TI on a uniform and conventional $s$-wave SC, in sharp
contrast to previous proposals of realizing MKPs of corner modes \citep{WangPRL2018,YanPRL2018,VolpezPRL2019,ZengPRL2019,Laubscher2020}.
Again, this result does not require pseudospin to be fully inverted across the corner,
as long as no node exists for the pairing order on the auxiliary trajectories {[}dashed
lines in Fig. \ref{fig:1}(f,h){]}.

\begin{figure}[b]
\begin{centering}
\includegraphics{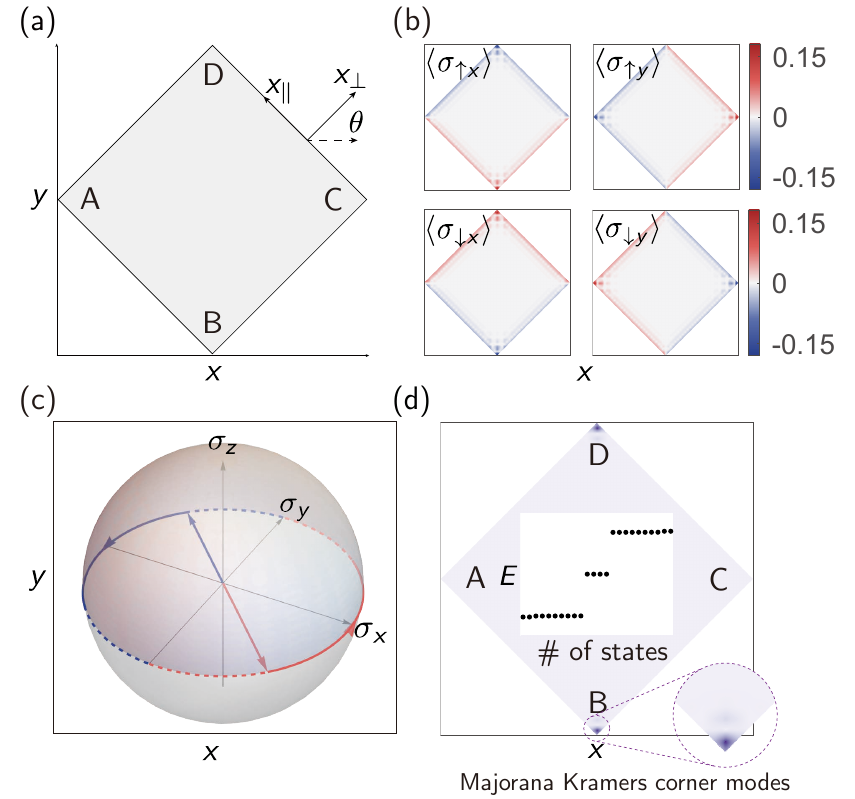}
\par\end{centering}
\caption{\label{fig:2}Pseudospin analysis for the first minimal model, with parameters $m=2,t=1,v=1$.
(a) Illustration of the sample geometry and normal vector of edges. (b) Numerical
results of the pseudospin polarizations of the two spin sectors. (c) Trajectories
of pseudospin polarizations of spin-up particle (red solid line) and spin-down particle
(blue solid line) states traveling from edge AB to BC, with valid continuation (dashed
lines). Concatenating two trajectories gives a complete winding which encircles $2\pi$
solid angle. (d) The distribution of the Dirac corner modes, and the energy level
plotted under magnetic SC pairing strength $\Delta_{s}=0.2$.}
\end{figure}

\begin{figure*}[t]
\begin{centering}
\includegraphics{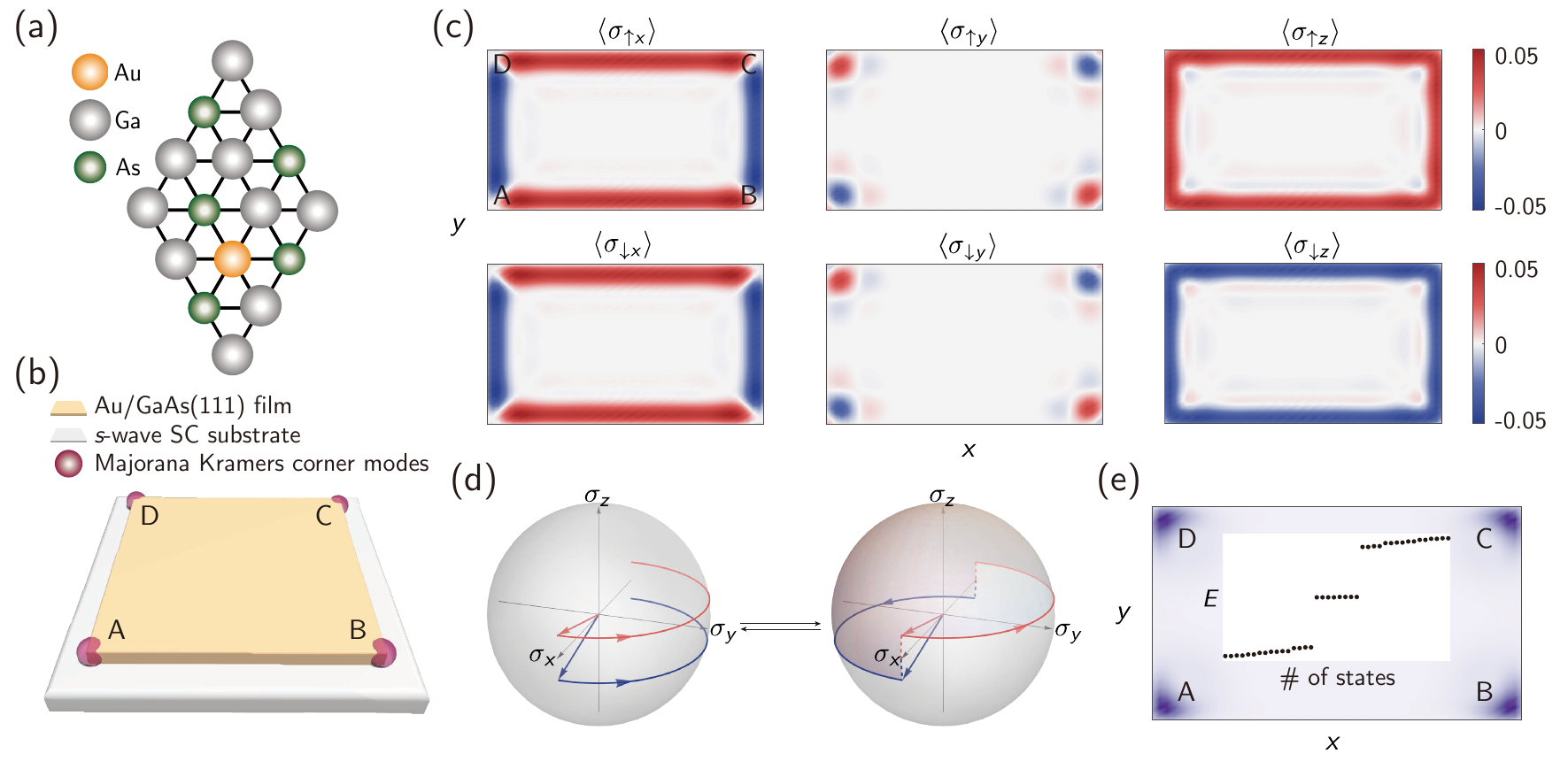}
\par\end{centering}
\caption{\label{fig:3}The material realization of SOTSC and its pseudospin analysis. (a)
$\sqrt{3}\times\sqrt{3}R30^{\circ}$ trigonal superlattice of Au/GaAs(111) surface.
(b) Schematic plot of the QSH/$s$-wave SC heterostructure with MKPs of corner modes. (c) Numerical
results of the pseudospin polarizations. (d) Boundary pseudospin textures of the
2nd type QSH phase across corner B. Red and blue arcs are the real trajectories of
pseudospin polarizations while dashed lines denote the continuation trajectories.
(e) The distribution of MKPs and the energy levels. (c) and (e) are obtained with
parameters $\varepsilon_{s}=0.74$eV, $\varepsilon_{p}=0$eV, $t_{ss\sigma}=-0.04$eV,
$t_{sp\sigma}=0.04$eV, $t_{pp\sigma}=0.18$eV, $t_{pp\pi}=0.005$eV, $\lambda=0.06$eV,
$\mu=0.555$eV, $\Delta_{s}=0.05$eV.}
\end{figure*}

$\textit{Application to SOTSC}s$.\textemdash{}The edge two-particle Berry phase theory
provides an intuitive principle to realize SOTIs and SOTSCs, while we leave the discussion on SOTIs based on the first type of TI to the supplemental
material \citep{SM}, and hereby focus on the more nontrivial prediction of MKPs of corner modes. As required, we consider
a generic type-II TI with two orbitals $\alpha$ and $\beta$ transforming to each other
under TRS, and propose two minimal-model realizations with such TI proximately coupled
to an $s$-wave SC. The Hamiltonian in $\psi=(\psi_{\uparrow\alpha},\psi_{\uparrow\beta},\psi_{\downarrow\alpha},\psi_{\downarrow\beta})^{T}$
basis is $H_{1(2)}=\int\mathrm{d}\mathbf{k}\ \psi^{\dagger}(\mathbf{k})\mathcal{H}_{\text{TI-}1(2)}\psi(\mathbf{k})+H_{s\text{-wave}}$,
with $H_{s\text{-wave}}=\sum_{\sigma=\alpha,\beta}\int\mathrm{d}\mathbf{k}\ \Delta_{s}(\psi_{\uparrow\sigma}(\mathbf{k})\psi_{\downarrow\sigma}(-\mathbf{k})+\mathrm{h.c.)}$
being TR invariant, and the low-energy band Hamiltonian
\begin{equation}
\mathcal{H}_{\text{TI-}1(2)}=(m-t|\mathbf{k}|^{2})s_{z}\sigma_{z}+vk_{x}s_{0(z)}\sigma_{x}+vk_{y}s_{0(z)}\sigma_{y}.\label{eq:7-1}
\end{equation}

We consider square-shape boundary, with edges terminating the lattice in a tilted
way {[}Fig. \ref{fig:2}(a){]}. When $\operatorname{sgn}m\operatorname{sgn}t>0$,
for an arbitrary sample edge with normal vector $\hat{\mathbf{x}}_{\perp}=(\cos\theta,\sin\theta)$,
assuming a half-infinite sample area $x_{\perp}<0$, we can solve out the helical
edge states
\begin{equation}
|\chi_{\uparrow}\rangle_{1,2}=\begin{bmatrix}1\\
0
\end{bmatrix}_{s}\otimes\begin{bmatrix}\frac{1}{\sqrt{2}}\\
-\frac{ie^{i\theta}}{\sqrt{2}}
\end{bmatrix}_{\sigma},|\chi_{\downarrow}\rangle_{1,2}=\begin{bmatrix}0\\
1
\end{bmatrix}\otimes\begin{bmatrix}\frac{1}{\sqrt{2}}\\
\pm\frac{ie^{i\theta}}{\sqrt{2}}
\end{bmatrix}_{\sigma}\label{eq:6}
\end{equation}
Thus the pseudospin polarizations for $|\chi_{\uparrow,\downarrow}\rangle$ are $\langle\boldsymbol{\sigma}_{\uparrow}\rangle_{1,2}=(\sin\theta,-\cos\theta,0),\langle\boldsymbol{\sigma}_{\downarrow}\rangle_{1,2}=(\mp\sin\theta,\pm\cos\theta,0)$.
{[}Fig. \ref{fig:2}(b){]} We put here the numerical results for the first model, which is the case of the material candidate to be proposed later. The trajectories of two pseudospin polarizations
across corner B (similar for other corners) are depicted as solid lines in Fig. \ref{fig:2}(c),
with in-between paths untraveled. So we can make a valid continuation
{[}dashed lines in Fig. \ref{fig:2}(c){]} to concatenate the two trajectories, rendering
a two-particle $\pi$ Berry phase in the PP channel. According to the generic theory,
a mass domain wall arises at the corner on the real trajectories if applying an $s$-wave
SC, as shown in numerical results in Fig. \ref{fig:2}(d). Note that the model has mirror symmetries. In particular, the mirror-$x$ symmetry $\hat{M}_{x}=\mathrm{i}s_{x}\sigma_{y}\tau_{z}$ relates edges AB and BC, forcing a SC $\pi$-junction at intersection corner B, consistent with the Berry phase mechanism~\citep{SM}.

$\textit{Realistic materials}$.\textemdash Now we propose the material realization
of the above SOTSC. We focus on the Au/GaAs(111) thin film, which realizes QSH phase
in triangular lattice {[}Fig. \ref{fig:3}(a){]} based on first-principle calculation
\citep{WangNatCommun2016}. The orbital components without SOC around $\Gamma$ point
are described by the $sp^{2}$ basis $(s,p_{x}+ip_{y},p_{x}-ip_{y})$, which captures
the minimal model of the system. With SOC, two types of QSH phases are supported,
respectively corresponding to $s$-$p$ band inversion (the first type) and $p$-$p$
gap-opening (the second type). In the following, we will study the second type QSH
phase, whose effective Hamiltonian is obtained by eliminating the bottom $s$ orbital
from the original $sp^{2}$ orbitals \citep{WangNatCommun2016}. The Hamiltonian
of TI-SC heterostructure in $\psi=(p_{\uparrow+},p_{\uparrow-},p_{\downarrow+},p_{\downarrow-})^{T}$
basis is $H=\int\mathrm{d}\mathbf{k}\ \psi^{\dagger}(\mathbf{k})\mathcal{H}_{\mathrm{eff}}\psi(\mathbf{k})+H_{s\text{-wave}}$,
with $H_{s\text{-wave}}=\sum_{\sigma=+,-}\int\mathrm{d}\mathbf{k}\ \Delta_{s}(p_{\uparrow\sigma}(\mathbf{k})p_{\downarrow\sigma}(-\mathbf{k})+\mathrm{h.c.)}$
and
\begin{align}
\mathcal{H}_{\mathrm{eff}}(\mathbf{k})= & a_{0}s_{0}\sigma_{0}+a_{0}^{\prime}(k_{x}^{2}+k_{y}^{2})s_{0}\sigma_{0}+\lambda s_{z}\sigma_{z}\nonumber \\
 & +a_{x}(k_{x}^{2}-k_{y}^{2})s_{0}\sigma_{x}+a_{y}k_{x}k_{y}s_{0}\sigma_{y}.\label{eq:8}
\end{align}
As being required, here the TRS $\mathcal{T^{\prime\prime}}=is_{y}\sigma_{x}\mathcal{K}$
reverses both spin and orbitals \citep{SM}, and the necessary pseudospin texture
in generic theory is obtained. The pseudospin polarizations of an arbitrary edge
read $\langle\boldsymbol{\sigma}_{\uparrow}\rangle=(a_{\uparrow}\cos2\theta,a_{\uparrow}\sin2\theta,(1-a_{\uparrow}^{2})^{1/2}),\langle\boldsymbol{\sigma}_{\downarrow}\rangle=(a_{\downarrow}\cos2\theta,a_{\downarrow}\sin2\theta,-(1-a_{\downarrow}^{2})^{1/2})$.
It is easily verified that $a_{\uparrow}=a_{\downarrow}<0$ \citep{SM}, so pseudospin
texture turns out to satisfy $\left(\langle\sigma_{\uparrow x}\rangle,\langle\sigma_{\uparrow y}\rangle=\langle\sigma_{\downarrow x}\rangle,\langle\sigma_{\downarrow y}\rangle\right)$,
which gives the two-particle Berry phase in PP channel $\gamma_{pp}$. We further consider
a rectangular sample proximate to the conventional $s$-wave SC {[}Fig. \ref{fig:3}(b){]}.
When traveling across the corner B, the pseudospin trajectories {[}Fig. \ref{fig:3}(c){]}
are depicted on the Bloch sphere {[}Fig. \ref{fig:3}(d){]}. The real trajectories
of pseudospin polarizations of $|\chi_{\uparrow p}\rangle$ (red semicircle) and
$|\chi_{\downarrow p}\rangle$ (blue semicircle) can be connected by the continuation
(dashed lines) with no nodes of pairing order existing in the continuation trajectories.
Thus the two-particle $\pi$-Berry phase in PP channel is obtained. Consequently, an
intrinsic SC $\pi$-junction is resulted with a Majorana Kramers pair at the corner
B, similar for the corners A, C and D. The numerical results of MKPs and energy
spectrum {[}Fig. \ref{fig:3}(e){]} confirm our prediction. Also,
an effective mirror symmetry $\hat{M}_{\hat{n}}=-\mathrm{i}\tau_{z}s_{z}\sigma_{z}$ ($\hat{n}=\frac{\sqrt{2}}{2}\hat{x}+\frac{\sqrt{2}}{2}\hat{y}$) of the low-energy Hamiltonian can relate all adjacent edges, forcing the emergence of SC $\pi$ junctions on every corner \citep{SM}.

We note that our theory provides a generic guideline in search for realistic materials
of the SOTSCs with MKPs via a uniform $s$-wave pairing and second type of TIs. In particular, in the type-II TI system the TRS reverses both the spin part and the orbital degree of freedom. Consequently, the TIs formed by $p$-$p$ \citep{ZhouNanoLett2014,LiNanoscale2018},
$p$-$d$ \citep{QianScience2014,YangSciRep2014,HuangPRB2016}, $d$-$d$ \citep{WengPRB2015,SiNanoLett2016}
orbitals satisfying this condition are candidates to realize the SOTSCs from uniform
$s$-wave pairing.

$\textit{Discussion and conclusion}$.\textemdash{} We have uncovered
a new and fundamental mechanism by introducing a novel concept of two-particle edge Berry
phase to characterize and realize second-order TIs or TSCs. Unlike other basic mechanisms like Wilson loop and
topological multipole invariants in the bulk, the boundary two-particle Berry phase mechanism
provides a fundamental and intuitive principle which facilitates to uncover new high-order topological matter.
As concrete examples of the application, we have predicted an unprecedented result
that the Majorana Kramers pairs of corner modes are realized by directly coupling type-II TR invariant TI edge
to a uniform $s$-wave SC, which stands in sharp contrast to the previous proposals. We further predicted the Au/GaAs(111) film as a material candidate for realization of the SOTSC, and a generic class of candidate
materials are also discussed. Our prediction with new novel concept shall inspire further theoretical and experimental studies, including also the extension of the present study to higher dimensions.

$\textit{Acknowledgment}s$.\textemdash This work was supported by the National Key R\&D Program of China (Grant No. 2021YFA1400900), National Natural Science Foundation of China (No. 11825401 and 11921005), and the Strategic
Priority Research Program of Chinese Academy of Science (Grant No. XDB28000000).

\setcounter{equation}{0} \setcounter{figure}{0}
\setcounter{table}{0}
\renewcommand{\theparagraph}{\bf}
\renewcommand{\thefigure}{S\arabic{figure}}
\renewcommand{\theequation}{S\arabic{equation}}

\onecolumngrid
\flushbottom
\newpage

\newpage
\begin{center}
\large \textbf{Supplementary Material}
\end{center}
In this Supplemental Material, we derive the formalism for two-state Berry phases
of particle-hole channel and particle-particle channel in detail and make a complementary
discussion for the condition of the situation where no mass domain wall being induced
in a corner turning process. Besides, we provide the details on the pseudospin analysis
for the BHZ model and Kane-Mele model and the derivation of boundary pseudospin textures
of the Au/AsGa(111) film. We also provide the lattice symmetry analysis for the SOTIs
and SOTSCs.

\section{Formalism for two-state Berry phase}

\subsection{\label{subsec:Formalism-for-BerryA}Formalism for two-state Berry phase in HOTI:
particle-hole channel}

\begin{figure*}
\centering{}\includegraphics{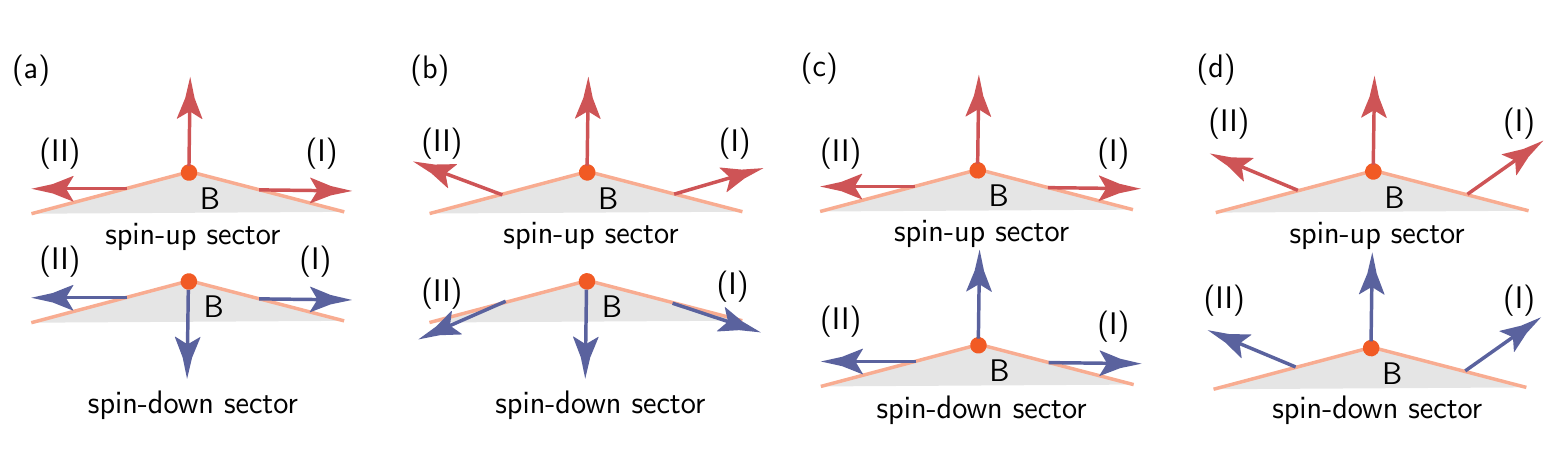}\caption{\label{fig:5}The pseudospin trajectories of both spin sectors across corner B. The
pseudospin trajectories of spin-up and spin-down sectors in (a) and (b) wind oppositely,
which allows a well-defined two-state Berry phase of particle-hole channel $\gamma_{ph}$.
While in the (c) and (d), the windings of pseudospin trajectories of both spin sectors
are in the same direction, leading to a vanished $\gamma_{ph}$ and a nontrivial
$\gamma_{pp}$.}
\end{figure*}

From an arbitrary edge $\mathrm{I}$ to its neighboring edge $\mathrm{II}$, the
evolution of spinors for pseudospin with the varying of boundary parameter $\theta^{(\mathrm{I})}\rightarrow\theta^{(\mathrm{II})}$
in the real space induces the evolution of spinors in the intrinsic space for pseudospin,
which can be represented by rotation on Bloch sphere for normalized spinors. Suppose
that the pseudospin vector of each spin sector always lies within $\sigma_{x}-\sigma_{y}$
plane of intrinsic space, its spinor can be generally represented as $|\chi_{s}(\alpha_{s})\rangle=e^{-i\frac{\alpha_{s}}{2}s_{0}(\sigma_{z}-\sigma_{0})}|\chi_{s}(0)\rangle\quad(s=\uparrow,\downarrow).$
The two spinors $|\chi_{s}(\alpha_{s})\rangle$ ($s=\uparrow,\downarrow$) are related
by TRS $|\chi_{\downarrow}(\alpha_{\downarrow})\rangle=\mathcal{T}|\chi_{\uparrow}(\alpha_{\uparrow})\rangle$.
For the case $\mathcal{T}=is_{y}\mathcal{K}$, the two pseudospin vectors wind along
opposite directions as $\theta$ varies, and we have the relation $\alpha_{\uparrow}=-\alpha_{\downarrow}$.
For spinors on edge I and II, we denote $|\chi_{s}^{(\mathrm{I},\mathrm{II})}\rangle=|\chi_{s}(\alpha_{s}^{(\mathrm{I},\mathrm{II})})\rangle$
and $\alpha_{s}^{(\mathrm{II})}-\alpha_{s}^{(\mathrm{I})}=\Delta\alpha_{s}$, so
$|\chi_{s}^{\mathrm{(II)}}\rangle=e^{-i\frac{\Delta\alpha_{s}}{2}s_{0}(\sigma_{z}-\sigma_{0})}|\chi_{s}^{\mathrm{(I)}}\rangle.$
Now consider the case where the pseudospin polarization of each spin sector gets
fully inverted when turning the corner, say, $\alpha_{\uparrow,\downarrow}^{(\mathrm{I})}=0$
and $\alpha_{\uparrow,\downarrow}^{(\mathrm{II})}=\pm\pi$. As shown in Fig. \ref{fig:5}(a),
since the pseudospin trajectory of each spin sector is non-closed, the Berry phase
for each particle state ($|\chi_{\uparrow}\rangle$ or $|\chi_{\downarrow}\rangle$)
can't be well defined. Interestingly, we find the trajectories of pseudospin of spin-up
sector and spin-down sector wind oppositely, which leads to that the joint pseudospin
trajectory of the particle state $|\chi_{\uparrow}\rangle$ and hole state $\overline{|\chi_{\downarrow}\rangle}$
(obtained by applying a particle-hole transformation on $|\chi_{\downarrow}\rangle$)
is closed (vice versa). Therefore it is reasonable to define a two-state Berry phase
in particle-hole channel:
\begin{equation}
\gamma_{ph}=\int_{\theta^{(\mathrm{I})}}^{\theta^{(\mathrm{II})}}\mathrm{d}\theta\ \langle\chi_{\uparrow}(\theta)|\otimes\overline{\langle\chi_{\downarrow}(\theta)|}(-i\partial_{\theta})|\chi_{\uparrow}(\theta)\rangle\otimes\overline{|\chi_{\downarrow}(\theta)\rangle}
\end{equation}
Since $\theta$ is solely determined by $\alpha_{\uparrow(\downarrow)}$ in each
spin sector, the formula can be reformulated as
\begin{align}
\gamma_{ph}= & \int_{\theta^{(\mathrm{I})}}^{\theta^{(\mathrm{II})}}\mathrm{d}\theta\ \langle\chi_{\uparrow}(\theta)|\otimes\overline{\langle\chi_{\downarrow}(\theta)|}(-i\partial_{\theta})\overline{|\chi_{\downarrow}(\theta)\rangle}\otimes|\chi_{\uparrow}(\theta)\rangle\nonumber \\
= & \int_{\theta^{(\mathrm{I})}}^{\theta^{(\mathrm{II})}}\mathrm{d}\theta\ \langle\chi_{\uparrow}(\theta)|(-i\partial_{\theta})|\chi_{\uparrow}(\theta)\rangle+\int_{\theta^{(\mathrm{I})}}^{\theta^{(\mathrm{II})}}\mathrm{d}\theta\ \overline{\langle\chi_{\downarrow}(\theta)|}(-i\partial_{\theta})\overline{|\chi_{\downarrow}(\theta)\rangle}\nonumber \\
= & \int_{0}^{\pi}\mathrm{d}\alpha_{\uparrow}\ \frac{\mathrm{d}\theta}{\mathrm{d}\alpha_{\uparrow}}\langle\chi_{\uparrow}(\theta(\alpha_{\uparrow}))|(-i\partial_{\theta})|\chi_{\uparrow}(\theta(\alpha_{\uparrow}))\rangle+\int_{0}^{-\pi}\mathrm{d}\alpha_{\downarrow}\ \frac{\mathrm{d}\theta}{\mathrm{d}\alpha_{\downarrow}}\langle\overline{\chi_{\downarrow}}(\theta(\alpha_{\downarrow}))|(-i\partial_{\theta})|\overline{\chi_{\downarrow}}(\theta(\alpha_{\downarrow}))\rangle\nonumber \\
= & \int_{0}^{\pi}\mathrm{d}\alpha_{\uparrow}\ \langle\chi_{\uparrow}(\theta(\alpha_{\uparrow}))|(-i\frac{\mathrm{d}\theta}{\mathrm{d}\alpha_{\uparrow}}\partial_{\theta})|\chi_{\uparrow}(\theta(\alpha_{\uparrow}))\rangle+\int_{0}^{-\pi}\mathrm{d}\alpha_{\downarrow}\ \overline{\langle\chi_{\downarrow}(\theta(\alpha_{\downarrow}))|}(-i\frac{\mathrm{d}\theta}{\mathrm{d}\alpha_{\downarrow}}\partial_{\theta})\overline{|\chi_{\downarrow}(\theta(\alpha_{\downarrow}))\rangle}\nonumber \\
= & \int_{0}^{\pi}\mathrm{d}\alpha_{\uparrow}\ \langle\chi_{\uparrow}(\alpha_{\uparrow})|(-i\partial_{\alpha_{\uparrow}})|\chi_{\uparrow}(\alpha_{\uparrow})\rangle+\int_{0}^{-\pi}\mathrm{d}\alpha_{\downarrow}\ \overline{\langle\chi_{\downarrow}(\alpha_{\downarrow})|}(-i\partial_{\alpha_{\downarrow}})\overline{|\chi_{\downarrow}(\alpha_{\downarrow})\rangle}\nonumber \\
= & \int_{0}^{\pi}\mathrm{d}\alpha_{\uparrow}\ \langle\chi_{\uparrow}(\alpha_{\uparrow})|(-i\partial_{\alpha_{\uparrow}})|\chi_{\uparrow}(\alpha_{\uparrow})\rangle-\int_{0}^{-\pi}\mathrm{d}\alpha_{\downarrow}\ \langle\chi_{\downarrow}(\alpha_{\downarrow})|(-i\partial_{\alpha_{\downarrow}})|\chi_{\downarrow}(\alpha_{\downarrow})\rangle
\end{align}
where $\overline{|\chi_{\downarrow}(\alpha_{\downarrow})\rangle}$ represents the
hole counterpart of $|\chi_{\downarrow}(\alpha_{\downarrow})\rangle$, which is obtained
by a particle-hole transformation $\overline{|\chi_{s}(\alpha_{s})\rangle}=e^{i\frac{\alpha_{s}}{2}s_{0}(\sigma_{z}-\sigma_{0})}\overline{|\chi_{s}(0)\rangle}$.
When introducing an uniform in-plane magnetic field $(\mathbf{M}\cdot\mathbf{s})\sigma_{0}$,
the mass term on an arbitrary edge is given by
\begin{align}
\mathcal{H}_{\mathrm{mass}}(\theta)= & \begin{bmatrix} & \langle\chi_{\uparrow}(\theta)|(\mathbf{M}\cdot\mathbf{s})\sigma_{0}|\chi_{\downarrow}(\theta)\rangle\\
\langle\chi_{\downarrow}(\theta)|(\mathbf{M}\cdot\mathbf{s})\sigma_{0}|\chi_{\uparrow}(\theta)\rangle
\end{bmatrix}
\end{align}
We can derive the differential equation for the matrix element,
\begin{align}
 & \mathrm{d}\langle\chi_{\uparrow}(\theta(\alpha_{\uparrow}))|(\mathbf{M}\cdot\mathbf{s})\sigma_{0}|\chi_{\downarrow}(\theta(\alpha_{\downarrow}))\rangle\nonumber \\
= & \mathrm{d}\alpha_{\uparrow}\langle\partial_{\alpha_{\uparrow}}\chi_{\uparrow}(\alpha_{\uparrow})|(\mathbf{M}\cdot\mathbf{s})\sigma_{0}|\chi_{\downarrow}(\alpha_{\downarrow})\rangle+\mathrm{d}\alpha_{\downarrow}\langle\chi_{\uparrow}(\alpha_{\uparrow})|(\mathbf{M}\cdot\mathbf{s})\sigma_{0}|\partial_{\alpha_{\downarrow}}\chi_{\downarrow}(\alpha_{\downarrow})\rangle\nonumber \\
= & \mathrm{d}\alpha_{\uparrow}\langle\partial_{\alpha_{\uparrow}}\chi_{\uparrow}(\alpha_{\uparrow})|\sum_{s}|\chi_{s}(\alpha_{s})\rangle\langle\chi_{s}(\alpha_{s})|(\mathbf{M}\cdot\mathbf{s})\sigma_{0}|\chi_{\downarrow}(\alpha_{\downarrow})\rangle+\mathrm{d}\alpha_{\downarrow}\langle\chi_{\uparrow}(\alpha_{\uparrow})|(\mathbf{M}\cdot\mathbf{s})\sigma_{0}\sum_{s}|\chi_{s}(\alpha_{s})\rangle\langle\chi_{s}(\alpha_{s})|\partial_{\alpha_{\downarrow}}\chi_{\downarrow}(\alpha_{\downarrow})\rangle\nonumber \\
= & (\mathrm{d}\alpha_{\uparrow}\langle\partial_{\alpha_{\uparrow}}\chi_{\uparrow}(\alpha_{\uparrow})|\chi_{\uparrow}(\alpha_{\uparrow})\rangle+\mathrm{d}\alpha_{\downarrow}\langle\chi_{\downarrow}(\alpha_{\downarrow})|\partial_{\alpha_{\downarrow}}\chi_{\downarrow}(\alpha_{\downarrow})\rangle)\langle\chi_{\uparrow}(\alpha_{\uparrow})|(\mathbf{M}\cdot\mathbf{s})\sigma_{0}|\chi_{\downarrow}(\alpha_{\downarrow})\rangle\nonumber \\
= & -i\left[\mathrm{d}\alpha_{\uparrow}\langle\chi_{\uparrow}(\alpha_{\uparrow})|(-i\partial_{\alpha_{\uparrow}})|\chi_{\uparrow}(\alpha_{\uparrow})\rangle+\mathrm{d}\alpha_{\downarrow}\overline{\langle\chi_{\downarrow}(\alpha_{\downarrow})}|(-i\partial_{\alpha_{\downarrow}})|\overline{\chi_{\downarrow}(\alpha_{\downarrow})\rangle}\right]\langle\chi_{\uparrow}(\alpha_{\uparrow})|(\mathbf{M}\cdot\mathbf{s})\sigma_{0}|\chi_{\downarrow}(\alpha_{\downarrow})\rangle,
\end{align}
where we have used the relation $\langle\chi_{\downarrow}(\alpha_{\downarrow})|\partial_{\alpha_{\downarrow}}\chi_{\downarrow}(\alpha_{\downarrow})\rangle=-\overline{\langle\chi_{\downarrow}(\alpha_{\downarrow})}|\partial_{\alpha_{\downarrow}}|\overline{\chi_{\downarrow}(\alpha_{\downarrow})\rangle}$.
Integrating from edge I to edge II, we obtain $\langle\chi_{\uparrow}^{(\mathrm{II})}|(\mathbf{M}\cdot\mathbf{s})\sigma_{0}|\chi_{\downarrow}^{(\mathrm{II})}\rangle=e^{-i(\int_{0}^{\pi}\mathrm{d}\alpha_{\uparrow}\ \langle\chi_{\uparrow}(\alpha_{\uparrow})|(-i\partial_{\alpha_{\uparrow}})|\chi_{\uparrow}(\alpha_{\uparrow})\rangle+\int_{0}^{-\pi}\mathrm{d}\alpha_{\downarrow}\ \overline{\langle\chi_{\downarrow}(\alpha_{\downarrow})|}(-i\partial_{\alpha_{\downarrow}})\overline{|\chi_{\downarrow}(\alpha_{\downarrow})\rangle})}\langle\chi_{\uparrow}^{(\mathrm{I})}|(\mathbf{M}\cdot\mathbf{s})\sigma_{0}|\chi_{\downarrow}^{(\mathrm{I})}\rangle=e^{-i\gamma_{ph}}\langle\chi_{\uparrow}^{(\mathrm{I})}|(\mathbf{M}\cdot\mathbf{s})\sigma_{0}|\chi_{\downarrow}^{(\mathrm{I})}\rangle$.
The particle-hole channel two-state Berry phase $\gamma_{ph}$ can be directly computed
as follows,
\begin{align}
\gamma_{ph}= & -i\int_{0}^{\pi}\mathrm{d}\alpha_{\uparrow}\langle\chi_{\uparrow}(\alpha_{\uparrow})|\partial_{\alpha_{\uparrow}}|\chi_{\uparrow}(\alpha_{\uparrow})\rangle-i\int_{0}^{-\pi}\mathrm{d}\alpha_{\downarrow}\overline{\langle\chi_{\downarrow}(\alpha_{\downarrow})|}\partial_{\alpha_{\downarrow}}\overline{|\chi_{\downarrow}(\alpha_{\downarrow})\rangle}\nonumber \\
= & -i\int_{0}^{\pi}\mathrm{d}\alpha_{\uparrow}\langle\chi_{\uparrow}(\alpha_{\uparrow})|\partial_{\alpha_{\uparrow}}|\chi_{\uparrow}(\alpha_{\uparrow})\rangle-i\int_{0}^{\pi}\mathrm{d}(-\alpha_{\downarrow})\overline{\langle\chi_{\downarrow}(-\alpha_{\downarrow})|}\partial_{(-\alpha_{\downarrow})}\overline{|\chi_{\downarrow}(-\alpha_{\downarrow})\rangle}\nonumber \\
= & -i\int_{0}^{\pi}\mathrm{d}\alpha^{\uparrow}\langle\chi_{\uparrow}(\alpha_{\uparrow})|\partial_{\alpha_{\uparrow}}|\chi_{\uparrow}(\alpha_{\uparrow})\rangle-i\int_{\pi}^{2\pi}\mathrm{d}\alpha_{\uparrow}\langle\chi_{\uparrow}(\alpha_{\uparrow})|\partial_{\alpha_{\uparrow}}|\chi_{\uparrow}(\alpha_{\uparrow})\rangle\nonumber \\
= & -i\int_{0}^{2\pi}\mathrm{d}\alpha_{\uparrow}\langle\chi_{\uparrow}(\alpha_{\uparrow})|\partial_{\alpha_{\uparrow}}|\chi_{\uparrow}(\alpha_{\uparrow})\rangle\nonumber \\
= & \ensuremath{\pm\pi\quad(\mathrm{mod\ }2\pi)},
\end{align}
 where we have used the relation $\alpha_{\uparrow}=-\alpha_{\downarrow}$. Now it
is readily seen that
\begin{align}
\langle\chi_{\uparrow}^{\mathrm{(II)}}|(\mathbf{M}\cdot\mathbf{s})\sigma_{0}|\chi_{\downarrow}^{\mathrm{\mathrm{(II)}}}\rangle= & e^{-i\gamma_{ph}}\langle\chi_{\uparrow}^{\mathrm{(I)}}|(\mathbf{M}\cdot\mathbf{s})\sigma_{0}|\chi_{\downarrow}^{\mathrm{\mathrm{(I)}}}\rangle=-\langle\chi_{\uparrow}^{\mathrm{(I)}}|(\mathbf{M}\cdot\mathbf{s})\sigma_{0}|\chi_{\downarrow}^{\mathrm{\mathrm{(I)}}}\rangle.
\end{align}

This result makes sure that a mass domain wall exists in the intersection of edge
I and II and enables a Dirac corner mode. In fact, the joint pseudospin trajectories
of $|\chi_{\uparrow}\rangle$ and $\mathrm{\overline{|\chi_{\downarrow}\rangle}}$
are usually non-closed {[}Fig. \ref{fig:5}(b){]}, but we can always connect them
by the way of continuation to obtain a closed loop on which the particle-hole channel
two-state Berry phase is well defined. And we need to determine whether the mass
term changes sign when traveling from edge I to edge II or not. We claim that there
must be a mass domain wall at corner B as long as one can find the continuation trajectories
in which no mass nodes exist.

\subsection{\label{subsec:Formalism-for-BerryB}Formalism for two-state Berry phase in HOTSC:
particle-particle channel }

Now we proceed to discuss the geometric phase relevant in the formation of SOTSC,
which is the two-state Berry phase in the particle-particle channel. Similarly, the
pseudospin vector is given by $|\chi_{s\tau}(\alpha_{s\tau})\rangle=e^{-i\frac{\alpha_{s\tau}}{2}\tau_{0}s_{0}(\sigma_{z}-\sigma_{0})}|\chi_{s\tau}(0)\rangle\quad(s=\uparrow,\downarrow;\tau=p,h)$.For
the case the TRS operator is modified and the two spinors wind along the same direction
as $\theta$ varies, we have $\mathrm{d}\alpha_{\uparrow p}(\theta)=\mathrm{d}\alpha_{\downarrow p}(\theta)$.
For simplicity, we take $\alpha_{\uparrow p}=\alpha_{\downarrow p}$, the discussion
for the cases where $\alpha_{\uparrow p}=\alpha_{\downarrow p}+\mathrm{const.}$
can be discussed similarly. For two adjacent edges I and II, we denote $|\chi_{sp}^{(\mathrm{I},\mathrm{II})}\rangle=|\chi_{sp}(\alpha_{sp}=\alpha_{s}^{(\mathrm{I},\mathrm{II})})\rangle$,
and we first consider the situation where the vectors of pseudospin get fully reversed
when traveling from edge I to edge II {[}Fig. \ref{fig:5}(c){]}, say, $\alpha_{\uparrow,\downarrow}^{(\mathrm{I})}=0$
and $\alpha_{\uparrow,\downarrow}^{(\mathrm{II})}=\pi$. Remarkably, since the two
pseudospins vectors within different spin sectors wind along the same direction across
the corner, the two-state Berry phase vanishes in the particle-hole channel, while
in a new channel (particle-particle channel) it comes into being. We define the two-state
Berry phase in the particle-particle channel as

\begin{align}
\gamma_{pp} & =\int_{\theta_{(\mathrm{I})}}^{\theta_{(\mathrm{II})}}\mathrm{d}\theta\ \langle\chi_{\uparrow p}(\theta)|\otimes\langle\chi_{\downarrow p}(\theta)|(-i\partial_{\theta})|\chi_{\uparrow p}(\theta)\rangle\otimes|\chi_{\downarrow p}(\theta))\rangle\nonumber \\
 & =\int_{\theta_{(\mathrm{I})}}^{\theta_{(\mathrm{II})}}\mathrm{d}\theta\ \langle\chi_{\uparrow p}(\theta)|(-i\partial_{\theta})|\chi_{\uparrow p}(\theta)\rangle+\int_{\theta_{(\mathrm{I})}}^{\theta_{(\mathrm{II})}}\mathrm{d}\theta\ \langle\chi_{\downarrow p}(\theta)|(-i\partial_{\theta})|\chi_{\downarrow p}(\theta))\rangle\nonumber \\
 & =-i\int_{\theta_{(\mathrm{I})}}^{\theta_{(\mathrm{II})}}\mathrm{d}\alpha_{\uparrow p}\frac{\mathrm{d}\theta(\alpha_{\uparrow p})}{\mathrm{d}\alpha_{\uparrow p}}\ \langle\chi_{\uparrow p}(\theta(\alpha_{\uparrow p}))|\partial_{\theta}|\chi_{\uparrow p}(\theta(\alpha_{\uparrow p}))\rangle-i\int_{\theta_{(\mathrm{I})}}^{\theta_{(\mathrm{II})}}\mathrm{d}\alpha_{\downarrow p}\frac{\mathrm{d}\theta(\alpha_{\downarrow p})}{\mathrm{d}\alpha_{\downarrow p}}\ \langle\chi_{\downarrow p}(\theta(\alpha_{\downarrow p}))|\partial_{\theta}|\chi_{\downarrow p}(\theta(\alpha_{\downarrow p}))\rangle\nonumber \\
 & =-i\int_{\theta_{(\mathrm{I})}}^{\theta_{(\mathrm{II})}}\mathrm{d}\alpha_{\uparrow p}\ \langle\chi_{\uparrow p}(\theta(\alpha_{\uparrow p}))|\frac{\mathrm{d}\theta(\alpha_{\uparrow p})}{\mathrm{d}\alpha_{\uparrow p}}\partial_{\theta}|\chi_{\uparrow p}(\theta(\alpha_{\uparrow p}))\rangle-i\int_{\theta_{(\mathrm{I})}}^{\theta_{(\mathrm{II})}}\mathrm{d}\alpha_{\downarrow p}\ \langle\chi_{\downarrow p}(\theta(\alpha_{\downarrow p}))|\frac{\mathrm{d}\theta(\alpha_{\downarrow p})}{\mathrm{d}\alpha_{\downarrow p}}\partial_{\theta}|\chi_{\downarrow p}(\theta(\alpha_{\downarrow p}))\rangle\nonumber \\
 & =\int_{0}^{\pi}\mathrm{d}\alpha_{\uparrow p}\ \langle\chi_{\uparrow p}(\alpha_{\uparrow p})|(-i\partial_{\alpha_{\uparrow p}})|\chi_{\uparrow p}(\alpha_{\uparrow p})\rangle+\int_{0}^{\pi}\mathrm{d}\alpha_{\downarrow p}\ \langle\chi_{\downarrow p}(\alpha_{\downarrow p})|(-i\partial_{\alpha_{\downarrow p}})|\chi_{\downarrow p}(\alpha_{\downarrow p})\rangle.
\end{align}
Since $|\chi_{sh}(\alpha_{sh})\rangle$ is obtained by a particle-hole transformation
on $|\chi_{sp}(\alpha_{sp})\rangle$,
\begin{align}
|\chi_{sh}(\alpha_{sh})\rangle= & \hat{\mathcal{C}}(|\chi_{sp}(\alpha_{sp})\rangle)\nonumber \\
= & \tau_{x}\mathcal{K}(e^{-i\frac{\alpha_{sp}}{2}\tau_{0}s_{0}(\sigma_{z}-\sigma_{0})}|\chi_{sp}^{\mathrm{(I)}}\rangle)\nonumber \\
= & e^{i\frac{\alpha_{sp}}{2}\tau_{0}s_{0}(\sigma_{z}-\sigma_{0})}|\chi_{sh}^{\mathrm{(I)}}\rangle,
\end{align}
and $|\chi_{sh}(\alpha_{sh})\rangle$ can also be written as $e^{-i\frac{\alpha_{sh}}{2}\tau_{0}s_{0}(\sigma_{z}-\sigma_{0})}|\chi_{sh}^{(\mathrm{I})}\rangle,$
we obtain the relation $\alpha_{sp}=-\alpha_{sh}.$ When introducing the conventional
$s$-wave SC pairing $-\Delta_{s}\tau_{y}s_{y}\sigma_{0}$, the mass term on an arbitrary
edge is given by
\begin{align}
\mathcal{H}_{\mathrm{mass}}(\theta)= & -\Delta_{s}\begin{bmatrix} &  &  & \langle\chi_{\uparrow p}(\theta)|\tau_{y}s_{y}\sigma_{0}|\chi_{\downarrow h}(\theta)\rangle\\
 &  & \langle\chi_{\downarrow p}(\theta)|\tau_{y}s_{y}\sigma_{0}|\chi_{\uparrow h}(\theta)\rangle\\
 & \langle\chi_{\uparrow h}(\theta)|\tau_{y}s_{y}\sigma_{0}|\chi_{\downarrow p}(\theta)\rangle\\
\langle\chi_{\downarrow h}(\theta)|\tau_{y}s_{y}\sigma_{0}|\chi_{\uparrow p}(\theta)\rangle
\end{bmatrix}.
\end{align}
Now the matrix element $\langle\chi_{\uparrow p}(\theta)|\tau_{y}s_{y}\sigma_{0}|\chi_{\downarrow h}(\theta)\rangle$
is not the edge particle-hole excitation gap as in the previous case, but is rather
the edge SC gap created through two-particle (Cooper) pairing. We derive the differential
equation for the matrix element
\begin{align}
 & \mathrm{d}\langle\chi_{\uparrow p}(\theta(\alpha_{\uparrow p}))|\tau_{y}s_{y}\sigma_{0}|\chi_{\downarrow h}(\theta(\alpha_{\downarrow h}))\rangle\nonumber \\
= & \mathrm{d}\alpha_{\uparrow p}\langle\partial_{\uparrow\alpha}\chi_{\uparrow p}(\alpha_{\uparrow p})|\tau_{y}s_{y}\sigma_{0}|\chi_{\downarrow h}(\alpha_{\downarrow h})\rangle+\mathrm{d}\alpha_{\downarrow h}\langle\chi_{\uparrow p}(\alpha_{\uparrow p})|\tau_{y}s_{y}\sigma_{0}|\partial_{\alpha_{\downarrow h}}\chi_{\downarrow h}(\alpha_{\downarrow h})\rangle\nonumber \\
= & \mathrm{d}\alpha_{\uparrow p}\langle\partial_{\uparrow\alpha}\chi_{\uparrow p}(\alpha_{\uparrow p})|\sum_{s\tau}|\chi_{s\tau}(\alpha_{s\tau})\rangle\langle\chi_{s\tau}(\alpha_{s\tau})|\tau_{y}s_{y}\sigma_{0}|\chi_{\downarrow h}(\alpha_{\downarrow h})\rangle+\mathrm{d}\alpha_{\downarrow h}\langle\chi_{\uparrow p}(\alpha_{\uparrow p})|\tau_{y}s_{y}\sigma_{0}\sum_{s\tau}|\chi_{s\tau}(\alpha_{s\tau})\rangle\langle\chi_{s\tau}(\alpha_{s\tau})|\partial_{\alpha_{\downarrow h}}\chi_{\downarrow h}(\alpha_{\downarrow h})\rangle\nonumber \\
= & (\mathrm{d}\alpha_{\uparrow p}\langle\partial_{\uparrow\alpha}\chi_{\uparrow p}(\alpha_{\uparrow p})|\chi_{\uparrow p}(\alpha_{s\tau})+\mathrm{d}\alpha_{\downarrow h}\langle\chi_{\downarrow h}(\alpha_{\downarrow h})|\partial_{\alpha_{\downarrow h}}\chi_{\downarrow h}(\alpha_{\downarrow h})\rangle)\langle\chi_{s\tau}(\alpha_{s\tau})|\tau_{y}s_{y}\sigma_{0}|\chi_{\downarrow h}(\alpha_{\downarrow h})\rangle\nonumber \\
= & -i\left[(\mathrm{d}\alpha_{\uparrow p}\langle\chi_{\uparrow p}(\alpha_{\uparrow p})|(-i\partial_{\alpha_{\uparrow p}})|\chi_{\uparrow p}(\alpha_{\uparrow p})\rangle+\mathrm{d}\alpha_{\downarrow p}\langle\chi_{\downarrow p}(\alpha_{\downarrow p})|(-i\partial_{\alpha_{\downarrow p}})|\chi_{\downarrow p}(\alpha_{\downarrow p})\rangle)\right]\langle\chi_{\uparrow p}(\alpha_{\uparrow p})|\tau_{y}s_{y}\sigma_{0}|\chi_{\downarrow h}(\alpha_{\downarrow h})\rangle\rangle,
\end{align}
where we have used the relation $\langle\chi_{\downarrow h}(\alpha_{\downarrow h})|\partial_{\alpha_{\downarrow h}}\chi_{\downarrow h}(\alpha_{\downarrow h})\rangle=\langle\chi_{\downarrow p}(\alpha_{\downarrow p})|\partial_{\alpha_{\downarrow p}}\chi_{\downarrow p}(\alpha_{\downarrow p})\rangle$
and $\alpha_{\downarrow p}=-\alpha_{\downarrow h}$. So, integrating from edge I
to edge II, we get $\langle\chi_{\uparrow p}^{\mathrm{(II)}}|\tau_{y}s_{y}\sigma_{0}|\chi_{\downarrow h}^{\mathrm{(II)}}\rangle=e^{-i(\int_{0}^{\pi}\mathrm{d}\alpha_{\uparrow p}\ \langle\chi_{\uparrow p}(\alpha_{\uparrow p})|(-i\partial_{\alpha_{\uparrow p}})|\chi_{\uparrow p}(\alpha_{\uparrow p})\rangle+\int_{0}^{\pi}\mathrm{d}\alpha_{\downarrow p}\ \langle\chi_{\downarrow p}(\alpha_{\downarrow p})|(-i\partial_{\alpha_{\downarrow p}})|\chi_{\downarrow p}(\alpha_{\downarrow p})\rangle)}\langle\chi_{\uparrow p}^{\mathrm{(I)}}|\tau_{y}s_{y}\sigma_{0}|\chi_{\downarrow h}^{\mathrm{(I)}}\rangle=e^{-i\gamma_{pp}}\langle\chi_{\uparrow p}^{\mathrm{(I)}}|\tau_{y}s_{y}\sigma_{0}|\chi_{\downarrow h}^{\mathrm{(I)}}\rangle$.
The two-state particle-particle channel Berry phase $\gamma_{pp}$ can be calculated
straightforwardly,
\begin{align}
\gamma_{pp}= & -i\int_{0}^{\pi}\mathrm{d}\alpha_{\uparrow p}\ \langle\chi_{\uparrow p}(\alpha_{\uparrow p})|\partial_{\alpha_{\uparrow p}}|\chi_{\uparrow p}(\alpha_{\uparrow p})\rangle-i\int_{0}^{\pi}\mathrm{d}\alpha_{\downarrow p}\ \langle\chi_{\downarrow p}(\alpha_{\downarrow p})|\partial_{\alpha_{\downarrow p}}|\chi_{\downarrow p}(\alpha_{\downarrow p})\rangle\nonumber \\
= & -i\int_{0}^{\pi}\mathrm{d}\alpha_{\uparrow p}\ \langle\chi_{\uparrow p}(\alpha_{\uparrow p})|\partial_{\alpha_{\uparrow p}}|\chi_{\uparrow p}(\alpha_{\uparrow p})\rangle-i\int_{\pi}^{2\pi}\mathrm{d}\alpha_{\downarrow p}\ \langle\chi_{\downarrow p}(\alpha_{\downarrow p})|\partial_{\alpha_{\downarrow p}}|\chi_{\downarrow p}(\alpha_{\downarrow p})\rangle\nonumber \\
= & -i\int_{0}^{\pi}\mathrm{d}\alpha_{\uparrow p}\ \langle\chi_{\uparrow p}(\alpha_{\uparrow p})|\partial_{\alpha_{\uparrow p}}|\chi_{\uparrow p}(\alpha_{\uparrow p})\rangle-i\int_{\pi}^{2\pi}\mathrm{d}\alpha_{\uparrow p}\ \langle\chi_{\uparrow p}(\alpha_{\uparrow p})|\partial_{\alpha_{\uparrow p}}|\chi_{\uparrow p}(\alpha_{\uparrow p})\rangle\nonumber \\
= & -i\int_{0}^{2\pi}\mathrm{d}\alpha_{\uparrow p}\ \langle\chi_{\uparrow p}(\alpha_{\uparrow p})|\partial_{\alpha_{\uparrow p}}|\chi_{\uparrow p}(\alpha_{\uparrow p})\rangle\nonumber \\
= & \ensuremath{\pm\pi\quad(\mathrm{mod\ }2\pi)}.
\end{align}
where we have used the relation $|\chi_{\downarrow}(\alpha_{\downarrow})\rangle=\mathcal{\mathcal{T}^{\prime\prime}}|\chi_{\uparrow}(\alpha_{\uparrow})\rangle$
and $\alpha_{\uparrow p}=\alpha_{\downarrow p}$. Now it is readily seen that
\begin{align}
\langle\chi_{\uparrow p}^{\mathrm{(II)}}|\tau_{y}s_{y}\sigma_{0}|\chi_{\downarrow h}^{\mathrm{(II)}})\rangle= & e^{-i\gamma_{pp}}\langle\chi_{\uparrow p}^{\mathrm{(I)}}|\tau_{y}s_{y}\sigma_{0}|\chi_{\downarrow h}^{\mathrm{(I)}}\rangle=-\langle\chi_{\uparrow p}^{\mathrm{(I)}}|\tau_{y}s_{y}\sigma_{0}|\chi_{\downarrow h}^{\mathrm{(I)}}\rangle,
\end{align}
which allows the appearance of mass domain wall (intrinsic $\pi$ junction) at corner
B and implies a Majorana Kramers corner modes. The discussion in a more general case
{[}Fig. \ref{fig:5}(d){]} where the pseudospin polarization isn't fully inverted
across the corner (i.e. the joint trajectories of pseudospin of $|\chi_{\uparrow p}\rangle$
and $|\chi_{\downarrow p}\rangle$ are not closed) is the same as that in \ref{subsec:Formalism-for-BerryA}.

\section{\label{sec:Continuation-Condition-of}Continuation Condition of no mass domain wall }

\begin{figure}[h]
\begin{centering}
\includegraphics{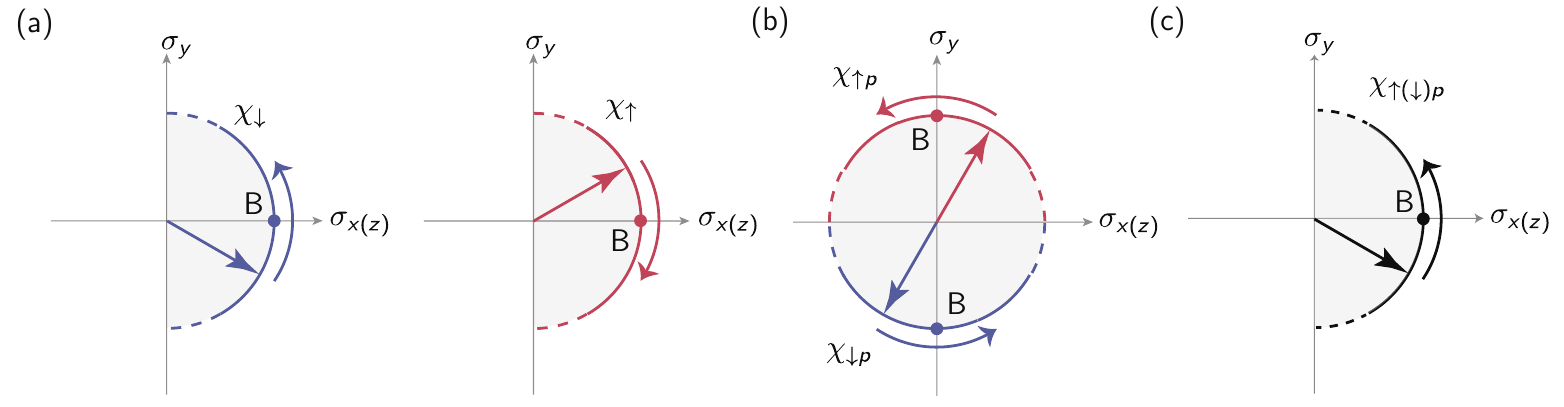}
\par\end{centering}
\caption{\label{fig:6} The pseudospin texture in (a) can't guarantee the mass domain wall
at corner under a uniform in-plane magnetic field. Solid lines represent the real
trajectories while the dashed lines denote the auxiliary paths required by the continuation.
Similarly, the pseudospin trajectories in (b) and (c) fail to induce the Majorana
Kramers corner modes. The straight red and blue arrows denote the pseudospin vectors
for $|\chi_{\uparrow}\rangle$ and $|\chi_{\downarrow}\rangle$, respectively and
the straight black arrow represents the coinciding of the two pseudospin vectors.}
\end{figure}

In this section, we will elucidate when exactly which continuations should be used
and discuss the situation where the mass nodes exist on the continuation trajectories
when the pseudospin polarization of each spin sector isn't fully reversed across
the corner. We have shown that the two-state Berry phase $\gamma_{ph}$($\gamma_{pp}$)
is well defined when the joint pseudospin trajectory of particle state $|\chi_{\uparrow}\rangle$
and hole state $\overline{|\chi_{\downarrow}\rangle}$ (particle state $|\chi_{\uparrow p}\rangle$
and particle state $|\chi_{\downarrow p}\rangle$) traveling from edge I to II is
closed {[}Fig. 1(c,e,g) in the main text{]}. Furthermore, this correspondence can
be generalized to a more generic case where the joint pseudospin trajectories are
non-closed {[}Fig. 1(d,f,h){]}. Hereafter we show how we can find the continuation
trajectories in this situation. We take the case in SOTIs for example. Assume that
pseudospins of $|\chi_{\uparrow,\downarrow}\rangle$ are confined to $\sigma_{x}-\sigma_{y}$
plane, the spin-helical edge states are related by the time-reversal symmetry, so
pseudospin textures satisfy $\left(\langle\sigma_{\uparrow x}\rangle,\langle\sigma_{\uparrow y}\rangle\right)=\left(\langle\sigma_{\downarrow x}\rangle,-\langle\sigma_{\downarrow y}\rangle\right)$.
Since the pseudospin of each spin sector isn't fully reversed across the corner {[}solid
lines in Fig. 1(d){]}, the two-state Berry phase in PH channel is not well defined
unless we consider the continuation trajectories {[}dashed lines in Fig. 1(d){]},
by which the pseudospins get fully inverted when traveling from edge I to II. Remarkably,
the pseudospins in the continuation trajectories are also constrained by time-reversal
symmetry. And if we can't find mass node on this auxiliary trajectories, the corner
must be mass domain wall.

What will happen when traveling across a corner if mass nodes appear in the continuation
trajectories? It is notable that the pseudospin vectors of spin-up state and spin-down
state should polarize oppositely in order to obtain a mass node under a uniform in-plane
Zeeman field or a conventional $s$-wave pairing. As shown in Fig. \ref{fig:6}(a,b,c),
the pseudospin vectors of $|\chi_{\uparrow}\rangle$ and $|\chi_{\downarrow}\rangle$
($|\chi_{\uparrow p}\rangle$ and $|\chi_{\downarrow h}\rangle$) can polarize oppositely
in the dashed lines (continuation trajectories) while the mass terms are always non-zero
and don't change signs in the real trajectories. Therefore, this type of pseudospin
textures does not guarantee a mass domain wall at corner and no in-gap corner modes
will appear in this special case.

Above consideration can be applied to the concrete models discussed in the main text.
For instance, since the pseudospin trajectories around corner B and D in Fig. 2(d)
behave like that in Fig. \ref{fig:6}(a) and we can only identify the pseudospin
vectors polarizing oppositely in the continuation paths, the pseudospin textures
does not lead to a mass domain wall at each corner under magnetic field. As a result,
one can't observe the appearance of corner modes at B and D. Numerical results shown
in Fig. 2(c) confirm our predictions. Similarly, we can also apply this analysis
on the corners which can't host the Majorana Kramers corner modes under a conventional
$s$-wave order {[}Fig. 2{]}. These examples reflect the validity of our edge geometric
phase mechanism.

\section{Pseudospin Analysis for the second-order topological insulators}

The two prominent models for 2D TIs, the Bernevig-Hughes-Zhang (BHZ) model \citep{NiuQianPRL2020}
and the Kane-Mele model \citep{SAYangPRL2020}, both support SOTI phase when subject
to an in-plane Zeeman field. We will show the application of edge geometric phase
mechanism to these models in detail. Remarkably, the corner modes are pinned at zero
energy in the presence of chiral symmetry, while they will be pushed away from zero
energy when breaking chiral symmetry. For the sake of simplicity, we take the BHZ
model for example to demonstrate this situation. And the Kane-Mele model is assumed
to preserve the chiral symmetry.

\subsection{Application to Bernevig-Hughes-Zhang (BHZ) model}

\begin{figure}[h]
\centering{}\includegraphics{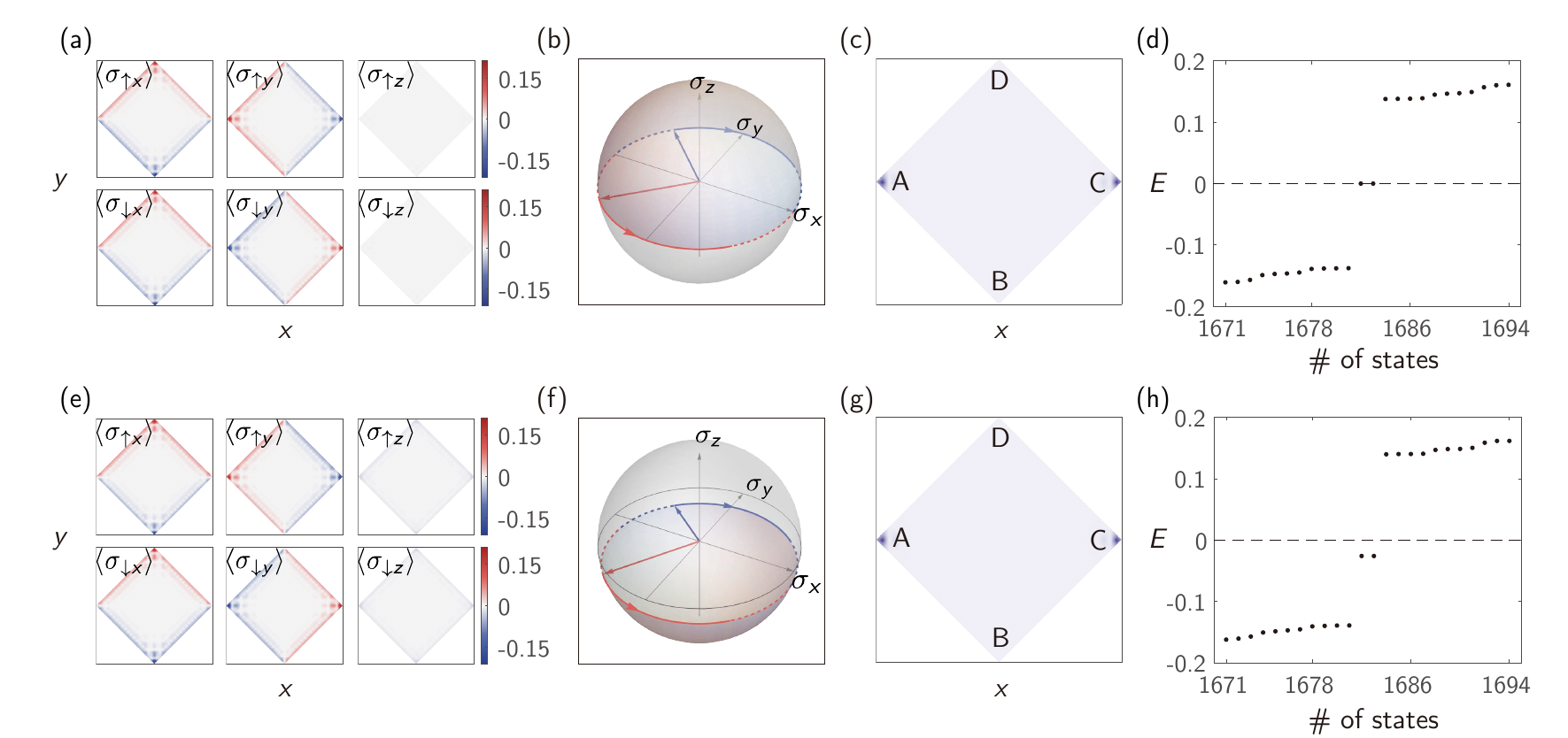}\caption{\label{fig:Figure 6}Schematics of the pseudospin analysis for the BHZ model, with
parameters $E=0.3,t_{0}=0.15,m=2,t=1,v=1$. We first show the results in the presence
of chiral symmetry. (a) Numerical results of the pseudospin polarizations of the
two spin sectors. (b) Trajectories of pseudospin polarizations of spin-up particle
(red solid line) and spin-down particle (blue solid line) states traveling from edge
BC to CD, with valid continuation (dashed lines). Reversing the direction for the
spin-down trajectory gives a complete winding which encircles $2\pi$ solid angle.
(c) and (d) The distribution of the Dirac corner modes, and the energy level plotted
under magnetic field strength $|\mathbf{M}|=0.2$. As for the case without chiral
symmetry, one can refer to (e), (f), (g) and (h).}
\end{figure}

We elucidate that the corner modes originate from the edge geometric phase in PH
channel for the BHZ model, which is similar for the Kane-Mele model. The Hamiltonian
is $\mathcal{H}(\mathbf{k})=\mathcal{H}_{\mathrm{BHZ}}(\mathbf{k})+\mathcal{H}_{\mathrm{Z}}$,
with $\mathcal{H}_{\mathrm{Z}}=M_{x}s_{x}+M_{y}s_{y}$ and
\begin{equation}
\mathcal{H}_{\mathrm{BHZ}}(\mathbf{k})=(E-t_{0}|\mathbf{k}|^{2})s_{0}\sigma_{0}+(m-t|\mathbf{k}|^{2})s_{0}\sigma_{z}+vk_{x}s_{z}\sigma_{x}+vk_{y}s_{0}\sigma_{y}.\label{eq:5}
\end{equation}
We consider square-shape boundary, with edges terminating the lattice in a tilted
way {[}Fig. 2(a){]}. When $\operatorname{sgn}m\operatorname{sgn}t>0$, for an arbitrary
sample edge with normal direction $\hat{\mathbf{x}}_{\perp}=(\cos\theta,\sin\theta)$,
assuming a half-infinite sample area $x_{\perp}<0$, the effective Hamiltonian can
be written as
\begin{align}
\mathcal{H}_{\mathrm{BHZ}}(\mathbf{k}) & =[E-t_{0}(k_{\parallel}^{2}+k_{\perp}^{2})]s_{0}\sigma_{0}+[m-t(k_{\parallel}^{2}+k_{\perp}^{2})]s_{0}\sigma_{z}\nonumber \\
 & +v(\sin\theta s_{z}\sigma_{x}-\cos\theta s_{0}\sigma_{y})k_{\parallel}+v(\cos\theta s_{z}\sigma_{x}+\sin\theta s_{0}\sigma_{y})k_{\perp}.
\end{align}
 $\mathcal{H}_{\mathrm{BHZ}}(\mathbf{k})$ can be separate into two parts: $\mathcal{H}_{\mathrm{0}}(k_{\perp})=(E-t_{0}k_{\perp}^{2})s_{0}\sigma_{0}+(m-tk_{\perp}^{2})s_{0}\sigma_{z}+v(\cos\theta s_{z}\sigma_{x}+\sin\theta s_{0}\sigma_{y})k_{\perp}$
and $\mathcal{H}_{\mathrm{p}}(k_{\parallel})=v(\sin\theta s_{z}\sigma_{x}-\cos\theta s_{0}\sigma_{y})k_{\parallel}$.
The edge states can be obtained by solving the equation $h_{0\uparrow,\downarrow}\psi_{\uparrow,\downarrow}(x_{\perp})=\epsilon\psi_{\uparrow,\downarrow}(x_{\perp})$:
with
\begin{align}
h_{0\uparrow,\downarrow} & =\begin{bmatrix}\begin{array}{cc}
(E-t_{0}k_{\perp}^{2})+(m-tk_{\perp}^{2}) & \pm e^{-\mathrm{i}\theta}vk_{\perp}\\
\pm e^{\mathrm{i}\theta}vk_{\perp} & (E-t_{0}k_{\perp}^{2})-(m-tk_{\perp}^{2})
\end{array}\end{bmatrix},
\end{align}
 We replace $k_{\perp}$ with $-\mathrm{i}\partial_{x_{\perp}}$ and take the trial
function $\psi_{\uparrow}(x_{\perp})=\begin{bmatrix}\begin{array}{c}
\alpha_{\uparrow}\\
\beta_{\uparrow}
\end{array}\end{bmatrix}e^{\lambda x_{\perp}}$ ($\lambda>0$). So the edge state for spin up takes the form of
\begin{equation}
\psi_{\uparrow}^{\epsilon}(x_{\perp})=\begin{bmatrix}\begin{array}{c}
\mathrm{i}e^{-\mathrm{i}\theta}v\lambda_{1}\\
(E-\epsilon+m)+(t+t_{0})\lambda_{1}^{2}
\end{array}\end{bmatrix}\mathcal{N}_{1}^{\uparrow}e^{\lambda_{1}x_{\perp}}+\begin{bmatrix}\begin{array}{c}
\mathrm{i}e^{-\mathrm{i}\theta}v\lambda_{2}\\
(E-\epsilon+m)+(t+t_{0})\lambda_{2}^{2}
\end{array}\end{bmatrix}\mathcal{N}_{2}^{\uparrow}e^{\lambda_{2}x_{\perp}}
\end{equation}
 with $\lambda_{1}>0$ and $\lambda_{2}>0$. The boundary condition at $x_{\perp}=0$
enforces the vanishing of the edge state wavefunction. In the end
\begin{equation}
\psi_{\uparrow}^{\epsilon}(x_{\perp})=(\mathcal{N}_{1}^{\uparrow}e^{\lambda_{1}x_{\perp}}+\mathcal{N}_{2}^{\uparrow}e^{\lambda_{2}x_{\perp}})\begin{bmatrix}\begin{array}{c}
\sqrt{\frac{t-t_{0}}{2t}}\\
-\mathrm{i}e^{\mathrm{i}\theta}\sqrt{\frac{t+t_{0}}{2t}}
\end{array}\end{bmatrix}.
\end{equation}
 Similarly, the spinor of edge state for the spin-down sector is $\chi_{\downarrow}=\begin{bmatrix}\begin{array}{c}
\sqrt{\frac{t-t_{0}}{2t}}\\
\mathrm{i}e^{-\mathrm{i}\theta}\sqrt{\frac{t+t_{0}}{2t}}
\end{array}\end{bmatrix}.$

Thus the pseudospin polarizations for $|\chi_{\uparrow,\downarrow}\rangle$ are $\langle\boldsymbol{\sigma}_{\uparrow,\downarrow}\rangle=(\frac{\sqrt{t^{2}-t_{0}^{2}}}{t}\sin\theta,\mp\frac{\sqrt{t^{2}-t_{0}^{2}}}{t}\cos\theta,-\frac{t_{0}}{t})$.
Consider the uniform in-plane Zeeman field $\mathcal{H}_{\mathrm{Z}}=M_{x}s_{x}+M_{y}s_{y}$
and project it onto the subspace spanned by the helical edge states, the effective
edge Hamiltonian is $\mathcal{H}_{\mathrm{edge}}=\frac{\sqrt{t^{2}-t_{0}^{2}}}{t}vk_{\parallel}\gamma_{z}+\sin\theta(-M_{x}\gamma_{y}+M_{y}\gamma_{x})-\frac{t_{0}}{t}\cos\theta(M_{x}\gamma_{x}+M_{y}\gamma_{y}).$
We first consider the case where $\mathcal{H}_{\mathrm{BHZ}}$ has a chiral symmetry,
which requires $t_{0}=0$ and leads to $\langle\boldsymbol{\sigma}_{\uparrow,\downarrow}\rangle=(\sin\theta,\mp\cos\theta,0)$
{[}Fig. \ref{fig:Figure 6}(a){]}. On the one hand, the trajectories of two pseudospin
polarizations across corner C (similar for other corners) are depicted as solid lines
in Fig. \ref{fig:Figure 6}(b), with in-between untraveled paths. We can make a valid
continuation {[}dashed lines in Fig. \ref{fig:Figure 6}(b){]} to concatenate the
two trajectories, rendering a two-state $\pi$ Berry phase in the PH channel. According
to the generic theory a mass domain arises at the corner if applying an in-plane
Zeeman field, as the mass term $\langle\chi_{\uparrow}|(\mathbf{M}\cdot\mathbf{s})\sigma_{0}|\chi_{\downarrow}\rangle$
is nonzero on the continuation paths, with numerical results shown in Fig. \ref{fig:Figure 6}(c,d).
On the other hand, the effective edge Hamiltonian reads $\mathcal{H}_{\mathrm{edge}}=vk_{\parallel}\gamma_{z}+\sin\theta(-M_{x}\gamma_{y}+M_{y}\gamma_{x})$.
which also has a chiral symmetry $\hat{\mathcal{S}}=\frac{M_{x}\gamma_{x}+M_{y}\gamma_{y}}{\sqrt{M_{x}^{2}+M_{y}^{2}}}$:
\begin{equation}
\hat{\mathcal{S}}^{-1}\mathcal{H}_{\mathrm{edge}}\hat{\mathcal{S}}=-\mathcal{H}_{\mathrm{edge}}.\label{eq:39}
\end{equation}
We have known that corner C serves as the mass domain wall which harbors zero-energy
corner state. Remarkably, the zero corner mode is also topologically protected due
to Eq. \ref{eq:39}.

We now consider the case with slight deviation from the chiral symmetry, where $t_{0}$
is non-zero but is relatively small compared to $t$ in realistic materials. The
pseudospin textures {[}Fig. \ref{fig:Figure 6}(e){]} $\langle\boldsymbol{\sigma}_{\uparrow,\downarrow}\rangle=(\frac{\sqrt{t^{2}-t_{0}^{2}}}{t}\sin\theta,\mp\frac{\sqrt{t^{2}-t_{0}^{2}}}{t}\cos\theta,-\frac{t_{0}}{t})$
around corner C are depicted in Fig. \ref{fig:Figure 6}(f). As we can see, the pseudospin
for $|\chi_{\uparrow,\downarrow}\rangle$ are not strictly constrained within $\sigma_{x}-\sigma_{y}$
plane, so the two-state Berry phase in PH channel only approximately equals to $\pi$.
Now the geometric phase doesn't necessitate a sign-change of the edge mass, thus
not guaranteeing a mass domain wall. The consequence is that the zero corner modes
are not pinned at zero energy, and can be lifted to be of finite energy, but still
within the gap {[}Fig. \ref{fig:Figure 6}(g) and (h){]}. This is justified from
the perspective of symmetry: as long as the symmetry-breaking term is not strong
enough to drive a phase transition from the HOTI phase to a trivial phase, the edge
effective Hamiltonian can be considered as having an approximate chiral symmetry:
\begin{equation}
\hat{\mathcal{S}}^{-1}\mathcal{H}_{\mathrm{edge}}\hat{\mathcal{S}}\approx-\mathcal{H}_{\mathrm{edge}}.
\end{equation}

When the chiral symmetry is broken, the fact that the two-state Berry phase can't
quantize to $\pi$ (even if we made proper continuation to make the pseudospin trajectories
closed) in turn illustrates that the edge geometric phase mechanism is complete for
understanding the behavior of corner modes.

\subsection{Application to Kane-Mele model}

The Kane-Mele model is defined on a honeycomb lattice with pseudospin degree of freedom
attributed to the two sublattices,
\begin{equation}
H=\sum_{\langle ij\rangle,\alpha}tc_{i\alpha}^{\dagger}c_{j\alpha}+\sum_{\langle\langle ij\rangle\rangle,\alpha\beta}it_{\mathrm{SO}}\nu_{ij}s_{\alpha\beta}^{z}c_{i\alpha}^{\dagger}c_{j\beta},
\end{equation}
where the first term is the nearest-neighbor hopping, and the second term is the
intrinsic SOC for the next-nearest-neighbor hopping, with $\nu_{ij}=+(-)$ if the
electron makes a left (right) turn during hopping from site $j$ to $i$. Different
from the models considered in the main text, where the band inversion momentum ($\Gamma$
point in the Brillouin zone) is coincident with a time-reversal invariant momentum
(TRIM), the Kane-Mele model has its band inversion momenta ($K,K^{\prime}$ point)
separate from the TRIM. We take the routine to solve the edge states at the momenta
where they are degenerate in energy (i.e.$\ $the TRIM), since in this way we can
directly diagnose the pseudospin polarization of the edge states. For this reason,
in the case of Kane-Mele model, the continuum $\mathbf{k}\cdot\mathbf{p}$ Hamiltonian
expanded around $K,K^{\prime}$ is troublesome to use. Instead, we adopt the lattice
approach, which essentially reduces to a quasi-one-dimensional eigenvalue problem
for each edge. As an example, we consider the corner-turning process traveling from
the lower edge (A) to the lower-right edge C, across the lower-right corner (B).

\begin{figure}[h]
\begin{centering}
\includegraphics{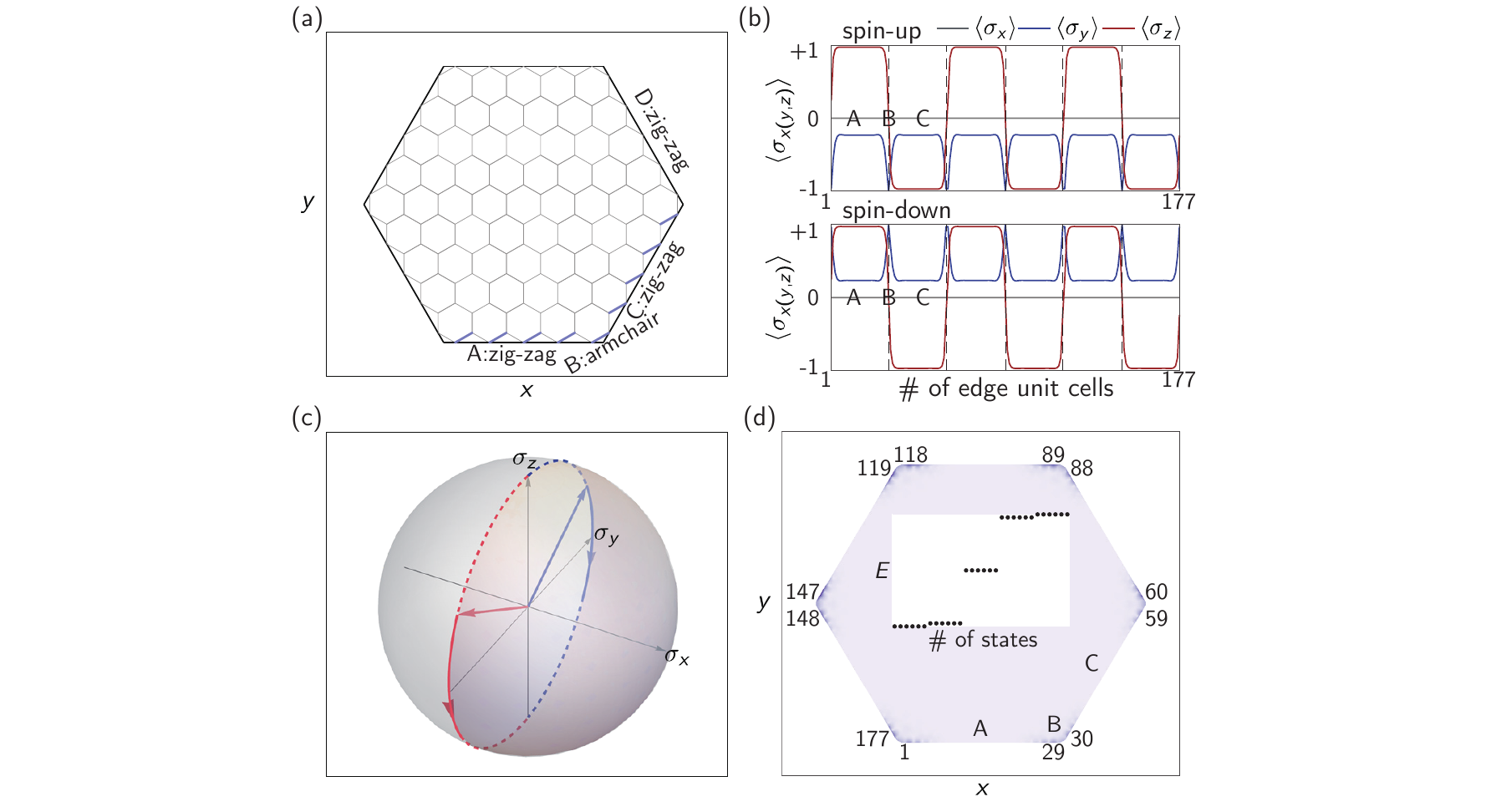}
\par\end{centering}
\caption{\label{fig:S2}Schematics of the pseudospin analysis for the Kane-Mele model, with
parameters $t=1,t_{\mathrm{so}}=0.06$. (a) Illustration of lattice configuration
and choice for the edge unit cells (colored links). The lower edge (A) and lower-right
edge (C) are of zig-zag type boundary condition, while the lower-right corner (B)
can be seen as being terminated from an armchair edge. (b) Distribution of the edge
states and the energy levels under magnetic field strength $|\mathbf{M}|=0.2$. (c)
Line plot of the pseudospin texture, with edge unit cell indices delineated in (b)
at each corner. (d) Trajectories of pseudospin polarizations of spin-up (red solid
line) and spin-down (blue solid line) edge states traveling from edge A to C. The
continuation (dashed lines) makes the joint trajectory close and encircle half the
Bloch sphere.}
\end{figure}

\subsubsection{Zig-zag edges}

The lower edge (A) is of zig-zag type boundary condition, and the way we choose the
edge unit cells is demonstrated as the colored links in Fig. \ref{fig:S2}(a). In
the lattice approach we need to solve the eigenvalue problem $H(k_{\parallel})|\Psi_{j}(k_{\parallel})\rangle=E(k_{\parallel})|\Psi_{j}(k_{\parallel})\rangle$,
here $\hat{e}_{\parallel}=\hat{e}_{x},\hat{e}_{\perp}=\hat{e}_{y}$. We denote $\mathbf{c}_{j}(k_{\parallel})=\left(c_{j,\uparrow,A}(k_{\parallel}),c_{j,\uparrow,B}(k_{\parallel}),c_{j,\downarrow,A}(k_{\parallel}),c_{j,\downarrow,B}(k_{\parallel})\right)^{T}$
and write the tight-binding Hamiltonian as
\begin{equation}
H(k_{\parallel})=\sum_{j}\mathbf{c}_{j}^{\dagger}(k_{\parallel})T_{0}(k_{\parallel})\mathbf{c}_{j}(k_{\parallel})+\mathbf{c}_{j+1}^{\dagger}(k_{\parallel})T_{+1}(k_{\parallel})\mathbf{c}_{j}(k_{\parallel})+\mathbf{c}_{j-1}^{\dagger}(k_{\parallel})T_{-1}(k_{\parallel})\mathbf{c}_{j}(k_{\parallel}),
\end{equation}
with $T_{0}(k_{\parallel})=t(1+\cos k_{\parallel})s_{0}\sigma_{x}+t\sin k_{\parallel}s_{0}\sigma_{y}-2t_{so}\sin k_{\parallel}s_{z}\sigma_{z}$,
$T_{+1}(k_{\parallel})=\frac{t}{2}e^{-i\frac{k_{\parallel}}{2}}s_{0}\sigma_{x}+\frac{it}{2}e^{-i\frac{k_{\parallel}}{2}}s_{0}\sigma_{y}+2t_{so}\sin\frac{k_{\parallel}}{2}s_{z}\sigma_{z}$
and $T_{-1}(k_{\parallel})=\frac{t}{2}e^{i\frac{k_{\parallel}}{2}}s_{0}\sigma_{x}-\frac{it}{2}e^{i\frac{k_{\parallel}}{2}}s_{0}\sigma_{y}+2t_{so}\sin\frac{k_{\parallel}}{2}s_{z}\sigma_{z}$,
and the ansatz $|\Psi_{j}(k_{\parallel})\rangle=\lambda^{j-1}|\Psi_{1}(k_{\parallel})\rangle\ (|\lambda|<1)$
plus the zig-zag boundary condition $|\Psi_{j=1}(k_{\parallel})\rangle=0$. The edge
states are degenerate in energy $E=0$ at TRIM $k_{\parallel}=\pi$, and the spinor
part of the eigenstates solved at this momentum are given by
\begin{equation}
|\chi_{\uparrow}\rangle^{(A)}=\frac{1}{\sqrt{1+\mu}}\begin{bmatrix}1\\
0
\end{bmatrix}_{s}\otimes\begin{bmatrix}1\\
-i\sqrt{\mu}
\end{bmatrix}_{\sigma},\quad|\chi_{\downarrow}\rangle^{(A)}=\frac{1}{\sqrt{1+\mu}}\begin{bmatrix}0\\
1
\end{bmatrix}_{s}\otimes\begin{bmatrix}1\\
i\sqrt{\mu}
\end{bmatrix}_{\sigma},
\end{equation}
with $\mu=\left(1+\frac{t^{2}}{8t_{so}^{2}}\right)-\sqrt{\left(1+\frac{t^{2}}{8t_{so}^{2}}\right)^{2}-1}$,
and corresponding pseudospin polarizations being $\langle\boldsymbol{\sigma}_{\uparrow,\downarrow}\rangle^{(A)}=\left(0,\mp\frac{2\sqrt{\mu}}{1+\mu},\frac{1-\mu}{1+\mu}\right)$.
In the same procedure as before but setting $\hat{e}_{\parallel}=\left(\frac{1}{2},\frac{\sqrt{3}}{2}\right),\hat{e}_{\perp}=\left(-\frac{\sqrt{3}}{2},\frac{1}{2}\right)$,
we can solve out the spinors of the edge states on the lower-right edge (C):
\begin{equation}
|\chi_{\uparrow}\rangle^{(C)}=\frac{1}{\sqrt{1+1/\mu}}\begin{bmatrix}1\\
0
\end{bmatrix}_{s}\otimes\begin{bmatrix}1\\
-\frac{i}{\sqrt{\mu}}
\end{bmatrix}_{\sigma},\quad|\chi_{\downarrow}\rangle^{(C)}=\frac{1}{\sqrt{1+1/\mu}}\begin{bmatrix}0\\
1
\end{bmatrix}_{s}\otimes\begin{bmatrix}1\\
\frac{i}{\sqrt{\mu}}
\end{bmatrix}_{\sigma}
\end{equation}
and read off their pseudospin polarizations: $\langle\boldsymbol{\sigma}_{\uparrow,\downarrow}\rangle^{(C)}=\left(0,\mp\frac{2\sqrt{\mu}}{1+\mu},-\frac{1-\mu}{1+\mu}\right)$.

\subsubsection{Armchair corner}

The lower-right corner (B) can be regarded as being terminated from an armchair edge,
and within the lattice approach its boundary condition is $|\Psi_{j=1}(k_{\parallel})\rangle=|\Psi_{j=2}(k_{\parallel})\rangle=0$.
In this case, $\hat{e}_{\parallel}=\hat{e}_{y},\hat{e}_{\perp}=-\hat{e}_{x}$, and
the spinors and pseudospin polarizations of the edge states turn out to be
\begin{equation}
|\chi_{\uparrow}\rangle^{(B)}=\begin{bmatrix}1\\
0
\end{bmatrix}_{s}\otimes\begin{bmatrix}1\\
-i
\end{bmatrix}_{\sigma},\ |\chi_{\downarrow}\rangle^{(B)}=\begin{bmatrix}0\\
1
\end{bmatrix}_{s}\otimes\begin{bmatrix}1\\
i
\end{bmatrix}_{\sigma};\qquad\langle\boldsymbol{\sigma}_{\uparrow}\rangle^{(B)}=(0,-1,0),\ \langle\boldsymbol{\sigma}_{\downarrow}\rangle^{(B)}=(0,1,0).
\end{equation}

\subsubsection{Pseudospin analysis}

The numerical results for the pseudospin of the two edge states are plotted in Fig.
\ref{fig:S2}(b), and their trajectories in the process traveling from edge A to
edge C across corner B are depicted as sold arcs in Fig. \ref{fig:S2}(c). In this
case, the pseudospin polarizations lie in the $\sigma_{y}-\sigma_{z}$ plane, but
the analysis is the same as that in the main text. It's easy to see that the two
spinors of pseudospin-$1/2$ always have a nonzero overlap when the pseudospin vectors
are restricted in the regions between the two solid segments, so a continuation can
be validly performed {[}dashed arcs in Fig. \ref{fig:S2}(c){]}, rendering a two-state
$\pi$ Berry phase in PH channel. After introducing the in-plane Zeeman field, a
mass domain wall must lie in the middle of the solid segments, corresponding to the
corner B. With the same analysis, the other five corners are all identified host
mass domain walls. The numerical results {[}Fig. \ref{fig:S2}(d){]} show exact six
corner states pinned to the six corners of the sample respectively, confirming our
pseudospin analysis.

\section{Derivation of Pseudospin textures of the second-type QSH phase of the material}

In this section, we focus on the second type of QSH phase of the material Au/GaAs(111).
As discussed in Ref. \citep{WangNatCommun2016}, effective Hamiltonian of its second
type of QSH phase around $\Gamma$ point in the basis $\psi=(p_{\uparrow+},p_{\uparrow-},p_{\downarrow+},p_{\downarrow-})^{T}$
can be written as
\begin{align}
\mathcal{H}_{\mathrm{eff}}(\mathbf{k})= & a_{0}s_{0}\sigma_{0}+a_{0}^{\prime}(k_{x}^{2}+k_{y}^{2})s_{0}\sigma_{0}+\lambda s_{z}\sigma_{z}\label{eq:32}\\
 & +a_{x}(k_{x}^{2}-k_{y}^{2})s_{0}\sigma_{x}+a_{y}k_{x}k_{y}s_{0}\sigma_{y},\nonumber
\end{align}
with the coefficients: $a_{0}=\varepsilon_{p}+3(t_{pp\pi}+t_{pp\sigma})$, $a_{0}^{\prime}=[-\frac{3}{4}(t_{pp\pi}+t_{pp\sigma})+\frac{9t_{sp\sigma}^{2}/2}{E-(\varepsilon_{s}+6t_{ss\sigma})}]$
and $a_{x}=\frac{a_{y}}{2}=[\frac{3}{8}(t_{pp\pi}-t_{pp\sigma})+\frac{9t_{sp\sigma}^{2}/2}{E-(\varepsilon_{s}+6t_{ss\sigma})}],$
where $\varepsilon_{s}$ and $\varepsilon_{p}$ are the on site energy for $s$ and
$p$ orbital, respectively. $E$ denote the chemical potential. $t_{ss\sigma}$,
$t_{sp\sigma}$, $t_{pp\sigma}$ and $t_{pp\pi}$ are N.N. hopping parameter, and
$\lambda$ is the SOC strength. Since the effective Hamiltonian in Eq. \ref{eq:32}
is written in the basis $(p_{x}+ip_{y},p_{x}-ip_{y})$ and the hybrid orbitals $p_{+}=p_{x}+ip_{y}$
and $p_{-}=p_{x}-ip_{y}$ are interchanged under time-reversal symmetry $\hat{\mathcal{T}}^{-1}p_{+}\hat{\mathcal{T}}=\hat{\mathcal{T}}^{-1}(p_{x}+ip_{y})\hat{\mathcal{T}}=p_{x}-ip_{y}=p_{-}$
. So $\hat{\mathcal{T}}$ should be modified to $\hat{\mathcal{T}^{\prime\prime}}=is_{y}\sigma_{x}\mathcal{K}$.
And we find that effective Hamiltonian $\mathcal{H}_{\mathrm{eff}}(\boldsymbol{k})$
of the second type of QSH is time-reversal invariant under this refined operator
\begin{equation}
\hat{\mathcal{T}^{\prime\prime}}^{-1}\mathcal{H}_{\mathrm{eff}}(\mathbf{k})\hat{\mathcal{T}^{\prime\prime}}=\mathcal{H}_{\mathrm{eff}}(-\mathbf{k}).
\end{equation}

We consider an arbitrary sample edge with normal direction $\hat{\mathbf{x}}_{\perp}=(\cos\theta,\sin\theta)$
and assume a half-infinite sample area $x_{\perp}<0$. The effective Hamiltonian
can be written as
\begin{align*}
\mathcal{H}_{\mathrm{eff}}(\mathbf{k}) & =a_{0}\sigma_{0}s_{0}+a_{0}^{\prime}(k_{\perp}^{2}+k_{\parallel}^{2})\sigma_{0}s_{0}+a_{x}[\cos2\theta(k_{\perp}^{2}-k_{\parallel}^{2})-2\sin2\theta k_{\perp}k_{\parallel}]\sigma_{x}s_{0}\\
 & +a_{y}[\frac{\sin2\theta}{2}(k_{\perp}^{2}-k_{\parallel}^{2})+\cos2\theta k_{\perp}k_{\parallel}]\sigma_{y}s_{0}+\lambda\sigma_{z}s_{z}.
\end{align*}

We replace $k_{\perp}\rightarrow-i\partial_{\perp}$ and decompose the Hamiltonian
as $\mathcal{H}_{\mathrm{eff}}=\mathcal{H}_{0}+\mathcal{H}_{p}$, where$\mathcal{H}_{0}(-i\partial_{\perp},k_{\parallel})=a_{0}\sigma_{0}s_{0}+a_{0}^{\prime}(-\partial_{\perp}^{2})\sigma_{0}s_{0}+\lambda\sigma_{z}s_{z}+a_{x}[\cos2\theta(-\partial_{\perp}^{2})]\sigma_{x}s_{0}+a_{y}[\frac{\sin2\theta}{2}(-\partial_{\perp}^{2})]\sigma_{y}s_{0}$
and $\mathcal{H}_{p}(-i\partial_{\perp},k_{\parallel})=-a_{x}2\sin2\theta k_{\parallel}(-i\partial_{\perp})\sigma_{x}s_{0}+a_{y}\cos2\theta k_{\parallel}(-i\partial_{\perp})\sigma_{y}s_{0}.$
Solving the eigenvalue equation $\mathcal{H}_{0}|\Psi_{\alpha}(x_{\perp})\rangle=\epsilon|\Psi_{\alpha}(x_{\perp})\rangle$
under the boundary conditions $|\Psi_{\alpha}(-\infty)\rangle=0$, we can obtain
two solutions at energy $\epsilon$. And the spinor parts of the two solutions are
\begin{equation}
|\chi_{\uparrow}\rangle=\frac{1}{\mathcal{N}_{\uparrow}}\left[\begin{array}{c}
1\\
0
\end{array}\right]_{s}\otimes\left[\begin{array}{c}
\frac{-\lambda a_{0}^{\prime}-\delta}{[\lambda+(a_{0}-\epsilon)]a_{x}}e^{-i2\theta}\\
1
\end{array}\right]_{\sigma},\quad|\chi_{\downarrow}\rangle=\frac{1}{\mathcal{N}_{\uparrow}}\left[\begin{array}{c}
0\\
1
\end{array}\right]_{s}\otimes\left[\begin{array}{c}
\frac{-\lambda a_{0}^{\prime}+\delta}{[\lambda-(a_{0}-\epsilon)]a_{x}}e^{-i2\theta}\\
1
\end{array}\right]_{\sigma},
\end{equation}
with normalization factors given by $\mathcal{N}_{\uparrow(\downarrow)}=\sqrt{1+\left(\frac{-\lambda a_{0}^{\prime}\mp\delta}{[\lambda\pm(a_{0}-\epsilon)]a_{x}}\right)^{2}}$,
where $\delta=\sqrt{\lambda^{2}a_{0}^{\prime2}-[\lambda^{2}-(a_{0}-\epsilon)^{2}]a_{x}^{2}}$.
We can set the chemical potential $E$ properly such that $\delta>0$. It is worth
noting that $|\chi_{\uparrow}\rangle$ and $|\chi_{\downarrow}\rangle$ can be related
by operator $\mathcal{T}^{\prime\prime}$: $is_{y}\sigma_{x}\mathcal{K}\left\{ \frac{1}{\mathcal{N}_{\uparrow}}\left[\begin{array}{c}
1\\
0
\end{array}\right]_{s}\otimes\left[\begin{array}{c}
\frac{-\lambda a_{0}^{\prime}-\delta}{[\lambda+(a_{0}-\epsilon)]a_{x}}e^{-i2\theta}\\
1
\end{array}\right]_{\sigma}\right\} =\frac{1}{\mathcal{N}_{\uparrow}}\left[\begin{array}{c}
0\\
1
\end{array}\right]_{s}\otimes\left[\begin{array}{c}
\frac{-\lambda a_{0}^{\prime}+\delta}{[\lambda-(a_{0}-\epsilon)]a_{x}}e^{-i2\theta}\\
1
\end{array}\right]_{\sigma}.$ Consequently, pseudospin polarizations of $|\chi_{\uparrow}\rangle$ and $|\chi_{\downarrow}\rangle$
can be expressed as $\langle\chi_{\uparrow,\downarrow}|\boldsymbol{\sigma}|\chi_{\uparrow,\downarrow}\rangle=\left(\begin{array}{ccc}
\frac{2}{\mathcal{N}_{\uparrow}^{2}}\frac{-\lambda a_{0}^{\prime}\mp\delta}{[\lambda\pm(a_{0}-\epsilon)]a_{x}}\cos2\theta & ,\frac{2}{\mathcal{N}_{\uparrow}^{2}}\frac{-\lambda a_{0}^{\prime}\mp\delta}{[\lambda\pm(a_{0}-\epsilon)]a_{x}}\sin2\theta, & \frac{1}{\mathcal{N}_{\uparrow}^{2}}\{\frac{-\lambda a_{0}^{\prime}\mp\delta}{[\lambda\pm(a_{0}-\epsilon)]a_{x}}\}^{2}-1\end{array}\right)$. We notice that $\frac{1}{\mathcal{N}_{\uparrow}^{2}}\frac{-\lambda a_{0}^{\prime}-\delta}{[\lambda+(a_{0}-\epsilon)]a_{x}}=\frac{1}{\mathcal{N}_{\downarrow}^{2}}\frac{-\lambda a_{0}^{\prime}+\delta}{[\lambda-(a_{0}-\epsilon)]a_{x}}$,
so $\langle\sigma_{\uparrow x}\rangle=\langle\sigma_{\downarrow x}\rangle,\quad\langle\sigma_{\uparrow y}\rangle=\langle\sigma_{\downarrow y}\rangle,\quad\langle\sigma_{\uparrow z}\rangle=-\langle\sigma_{\downarrow z}\rangle$.
This pseudospin textures indicate that the second type of QSH phase can support a
$\pi$ two-state Berry phase in PP channel, which is relevant to our scheme for realizing
SOTSC by pure $s$-wave SC.

In fact, the effective Hamiltonian can be recast into the BdG form:
\begin{align}
\mathcal{H}_{\mathrm{BdG}} & =\begin{bmatrix}\begin{array}{cc}
\mathcal{H}_{\mathrm{eff}}(\mathbf{k}) & 0\\
0 & -\mathcal{H}_{\mathrm{eff}}^{\ast}(-\mathbf{k})
\end{array}\end{bmatrix},\label{eq:35}
\end{align}
 and the particle-hole degree of freedom $\boldsymbol{\tau}$ will involve in the
spin-helical edge states:
\begin{align}
|\chi_{\uparrow p}\rangle=\frac{1}{\mathcal{N}_{\uparrow}}\left[\begin{array}{c}
1\\
0
\end{array}\right]_{\tau}\otimes\left[\begin{array}{c}
1\\
0
\end{array}\right]_{s}\otimes\left[\begin{array}{c}
\frac{-\lambda a_{0}^{\prime}-\delta}{[\lambda+(a_{0}-\epsilon)]a_{x}}e^{-i2\theta}\\
1
\end{array}\right]_{\sigma} & ;|\chi_{\downarrow p}\rangle=\frac{1}{\mathcal{N}_{\downarrow}}\left[\begin{array}{c}
1\\
0
\end{array}\right]_{\tau}\otimes\left[\begin{array}{c}
0\\
1
\end{array}\right]_{s}\otimes\left[\begin{array}{c}
\frac{-\lambda a_{0}^{\prime}+\delta}{[\lambda-(a_{0}-\epsilon)]a_{x}}e^{-i2\theta}\\
1
\end{array}\right]_{\sigma}\nonumber \\
|\chi_{\uparrow h}\rangle=\frac{1}{\mathcal{N}_{\uparrow}}\left[\begin{array}{c}
0\\
1
\end{array}\right]_{\tau}\otimes\left[\begin{array}{c}
1\\
0
\end{array}\right]_{s}\otimes\left[\begin{array}{c}
\frac{-\lambda a_{0}^{\prime}-\delta}{[\lambda+(a_{0}-\epsilon)]a_{x}}e^{i2\theta}\\
1
\end{array}\right]_{\sigma}; & |\chi_{\downarrow h}\rangle=\frac{1}{\mathcal{N}_{\downarrow}}\left[\begin{array}{c}
0\\
1
\end{array}\right]_{\tau}\otimes\left[\begin{array}{c}
0\\
1
\end{array}\right]_{s}\otimes\left[\begin{array}{c}
\frac{-\lambda a_{0}^{\prime}+\delta}{[\lambda-(a_{0}-\epsilon)]a_{x}}e^{i2\theta}\\
1
\end{array}\right]_{\sigma}.
\end{align}
Consequently, there are two pairs of spin-helical Majorana edge states compared with
the case of QSHI. Then we introduce the uniform $s$-wave SC $\mathcal{H}_{\mathrm{SC}}=-\Delta_{s}\tau_{y}s_{y}\sigma_{0}$
to gap out this gapless edge with the superconducting gap written as
\begin{align}
\mathcal{H}_{\mathrm{mass}} & =-\Delta_{s}\begin{bmatrix}\begin{array}{cccc}
0 & 0 & 0 & \langle\chi_{\uparrow p}|\tau_{y}s_{y}\sigma_{0}|\chi_{\downarrow h}\rangle\\
0 & 0 & \langle\chi_{\uparrow h}|\tau_{y}s_{y}\sigma_{0}|\chi_{\downarrow p}\rangle & 0\\
0 & \langle\chi_{\downarrow p}|\tau_{y}s_{y}\sigma_{0}|\chi_{\uparrow h}\rangle & 0 & 0\\
\langle\chi_{\downarrow h}|\tau_{y}s_{y}\sigma_{0}|\chi_{\uparrow p}\rangle & 0 & 0 & 0
\end{array}\end{bmatrix}\nonumber \\
 & =-\Delta_{s}\frac{\mathcal{N}_{\downarrow}}{\mathcal{N}_{\uparrow}}\frac{2}{\mathcal{N}_{\uparrow}^{2}}\frac{-\lambda a_{0}^{\prime}-\delta}{[\lambda+(a_{0}-\epsilon)]a_{x}}\cos2\theta s_{y}\tau_{y}=\Delta_{s}\frac{\mathcal{N}_{\downarrow}}{\mathcal{N}_{\uparrow}}\frac{2}{\mathcal{N}_{\downarrow}^{2}}\frac{-\lambda a_{0}^{\prime}+\delta}{[\lambda-(a_{0}-\epsilon)]a_{x}}\cos2\theta s_{y}\tau_{y}\nonumber \\
 & \propto-\Delta_{s}\cos2\theta s_{y}\tau_{y}.\label{eq:49}
\end{align}
It is easy to see that superconducting mass in Eq. \ref{eq:49} can change sign with
the variation of $\theta$, which indicates the formation of mass domain wall. Since
both the QSHI and $s$-wave SC are time-reversal invariant, Majorana Kramer pairs
will be bound at the mass domain wall corners. And because the bulk and edges of
the system are gapped, eventually we achieve the time-reversal invariant SOTSC. With
naturally imprinted particle-hole symmetry of the BdG formalism protecting the topology,
the Majorana Kramers pairs are pinned at zero energy without any other symmetry constraint
(e.g. chiral symmetry).

\section{\label{sec:Crystalline-symmetry-analysis}Crystalline symmetry analysis}

As stated in the main text, our edge geometric phase mechanism provides a consistent
and parallel approach to crystalline symmetry in identifying the higher-order topological
phases. We will take the examples in the main text and supplementary materials to
demonstrate the consistency in the following.

\paragraph{BHZ model}

The first example is the BHZ model discussed in the Sec. \ref{sec:Continuation-Condition-of}
with the Hamiltonian
\begin{equation}
\mathcal{H}_{\mathrm{BHZ}}(\boldsymbol{k})=(M+t\cos k_{x}+t\cos k_{y})s_{0}\sigma_{z}+v\sin k_{x}s_{z}\sigma_{x}+v\sin k_{y}s_{0}\sigma_{y}.
\end{equation}
 We have omitted the identity matrix term $(E-t_{0}|\mathbf{k}|^{2})s_{0}\sigma_{0}$
here for simplicity. Applying an uniform in-plane Zeeman field can drive this QSH
phase into a second-order topological insulator (SOTI) phase. As depicted in Fig.
\ref{fig:Figure 6}(c), the mirror symmetry $\hat{M_{y}}=-\mathrm{i}s_{x}\sigma_{z}$
is preserved by changing $y$ to $-y$ with the Zeeman field $\mathcal{H}_{\mathrm{Z}}=M_{x}s_{x}$
along the $x$ direction. And the total Hamiltonian $\mathcal{H}_{\mathrm{total}}=\mathcal{H}_{\mathrm{BHZ}}+\mathcal{H}_{\mathrm{Z}}$
satisfies $\hat{M_{y}}\mathcal{H}_{\mathrm{total}}(k_{x},k_{y})\hat{M_{y}}^{-1}=\mathcal{H}_{\mathrm{total}}(k_{x},-k_{y}).$
From the perspective of lattice symmetry, $\hat{M_{y}}=-\mathrm{i}s_{x}\sigma_{z}$
plays a central role in the second-order topological insulator phase, since it can
change the sign of Dirac mass terms of corresponding adjacent edges, which generates
a mass domain wall at the corner. In fact, the corner modes of SOTI phase can be
essentially attributed to the Jackiw-Rebbi soliton states, and the lattice symmetry
determines the positions of Dirac mass domain walls. For example, let's consider
an edge that preserves $\hat{M_{y}}$ and derive the corresponding spin-helical edge
states (spinor part) : $|\chi_{\uparrow}\rangle=\frac{1}{\sqrt{2}}\left[\begin{array}{c}
1\\
0
\end{array}\right]_{s}\otimes\begin{bmatrix}\begin{array}{c}
1\\
-\mathrm{i}
\end{array}\end{bmatrix}_{\sigma}$ and $|\chi_{\downarrow}\rangle=\frac{1}{\sqrt{2}}\left[\begin{array}{c}
0\\
1
\end{array}\right]_{s}\otimes\begin{bmatrix}\begin{array}{c}
1\\
\mathrm{i}
\end{array}\end{bmatrix}_{\sigma}$. Projecting the mirror symmetry operator $\hat{M_{y}}=-\mathrm{i}s_{x}\sigma_{z}$
onto the subspace spanned by $|\chi_{\uparrow}\rangle$ and $|\chi_{\downarrow}\rangle$
:
\begin{align}
\hat{\mathcal{M}_{y}} & =-\mathrm{i}\begin{bmatrix}\begin{array}{cc}
\langle\chi_{\uparrow}|s_{x}\sigma_{z}|\chi_{\uparrow}\rangle & \langle\chi_{\uparrow}|s_{x}\sigma_{z}|\chi_{\downarrow}\rangle\\
\langle\chi_{\downarrow}|s_{x}\sigma_{z}|\chi_{\uparrow}\rangle & \langle\chi_{\downarrow}|s_{x}\sigma_{z}|\chi_{\downarrow}\rangle
\end{array}\end{bmatrix}=-\mathrm{i}\gamma_{x}
\end{align}
 Then we concentrate on two adjacent edges ($\mathrm{BC}$,$\mathrm{CD}$) around
corner C . It is straightforward to see that edge $\mathrm{BC}$ and $\mathrm{CD}$
can be related by mirror symmetry $\hat{M_{y}}$ and their Dirac mass terms satisfy:
$\hat{\mathcal{M}_{y}}\mathrm{m}_{\mathrm{BC}}(\theta)|_{\theta=-\theta_{1}}\hat{\mathcal{M}_{y}}^{-1}=(-\mathrm{i}\gamma_{x})(M_{x}\sin\theta_{1}\gamma_{y})(\mathrm{i}\gamma_{x})=-M_{x}\sin\theta_{1}\gamma_{y}=\mathrm{m}_{\mathrm{CD}}(\theta)|_{\theta=\theta_{1}}$.
Consequently, mirror symmetry enforces the sign-changing between $\mathrm{m}_{\mathrm{BC}}(\theta)|_{\theta=-\theta_{1}}$
and $\mathrm{m}_{\mathrm{CD}}(\theta)|_{\theta=\theta_{1}}$, for which a topologically
protected 0D zero mode must exist at the mass domain wall corner C. Similarly, corner
A can also harbor a Dirac zero mode.

\paragraph{Kane-Mele model}

The tight-binding Hamiltonian of Kane-Mele model reads:
\begin{equation}
\mathcal{H}_{\mathrm{KM}}(\boldsymbol{k})=[h_{x}(\boldsymbol{k})\sigma_{x}+h_{y}(\boldsymbol{k})\sigma_{y}]s_{0}+h_{\mathrm{SO}}\sigma_{z}s_{z}+\lambda\sigma_{0}s_{y},
\end{equation}
 where $h_{x}(\boldsymbol{k})=t(1+2\cos3ak_{y}/2\cos\sqrt{3}ak_{x}/2)$, $h_{y}(\boldsymbol{k})=2t\sin3ak_{y}/2\cos\sqrt{3}ak_{x}/2$
and $h_{\mathrm{SO}}(\boldsymbol{k})=-2t_{\mathrm{SO}}(\sin\sqrt{3}ak_{x}-2\cos3ak_{y}\sin\sqrt{3}ak_{x}/2)$.
when the Zeeman field is along $y$ direction, the mirror symmetry $\hat{M}_{y}=\mathrm{i}\sigma_{x}s_{y}$
preserves by changing $y$ to $-y$. As shown in Fig. \ref{fig:S2}(a), considering
an $\hat{M}_{y}$-invariant edge ( Armchair edge ), the spin-helical edge states
are $|\chi_{\uparrow}\rangle=\frac{1}{\sqrt{2}}\begin{bmatrix}1\\
0
\end{bmatrix}_{s}\otimes\begin{bmatrix}1\\
-i
\end{bmatrix}_{\sigma}$ and $|\chi_{\downarrow}\rangle=\frac{1}{\sqrt{2}}\left[\begin{array}{c}
0\\
1
\end{array}\right]_{s}\otimes\begin{bmatrix}\begin{array}{c}
1\\
\mathrm{i}
\end{array}\end{bmatrix}_{\sigma}$ , so we can obtain the mirror symmetry operator projected in edge states subspace:
$\mathrm{i}\gamma_{x}$. Additionally, the Dirac mass terms of $\hat{M}_{y}$-related
edges C and D are $\mathrm{m}_{D}\gamma_{y}$ and $\mathrm{m}_{C}\gamma_{y}$. Since
$(\mathrm{i}\gamma_{x})\mathrm{m}_{C}\gamma_{y}(-\mathrm{i}\gamma_{x})=-\mathrm{m}_{C}\gamma_{y}=\mathrm{m}_{D}\gamma_{y}$,
a topologically protected 0D zero mode will be bound at the corner where edges C
and D intersect.

\paragraph{Toy model realizing MCKMs}

We only take the toy model (1) for example. The BdG Hamiltonian is
\begin{equation}
\mathcal{H}_{\mathrm{BdG}}^{(1)}=(M+t\cos k_{x}+t\cos k_{y})s_{z}\sigma_{z}\tau_{z}+v\sin k_{x}s_{0}\sigma_{x}\tau_{0}+v\sin k_{y}s_{0}\sigma_{y}\tau_{z}-\Delta_{s}s_{y}\sigma_{0}\tau_{y}
\end{equation}
 It is easy to see that $\hat{M}_{x}\mathcal{H}_{\mathrm{BdG}}^{(1)}(k_{x},k_{y})\hat{M}_{x}^{-1}=\mathcal{H}_{\mathrm{BdG}}^{(1)}(-k_{x},k_{y})$
, where mirror symmetry $\hat{M}_{x}=\mathrm{i}s_{x}\sigma_{y}\tau_{z}$. Considering
$\hat{M}_{x}$-related two edges $\mathrm{AB}$ and $\mathrm{BC}$ {[}Fig. 2(a){]},
we find that their SC mass terms have opposite signs due to the constraint of mirror
symmetry. As a result, corner B (mass domain wall) will bind the Majorana Kramers
 pair.

\paragraph{Realistic material Au/GaAs(111)}

Remarkably, it is rather hard to make sure about the lattice symmetries which are
preserved in this realistic system shown in Figure. 3, because we made the triangle
lattice fitted into a rectangle geometry. We can only say that the continuum Hamiltonian
$\mathcal{H}_{\mathrm{BdG}}$ {[}Eq. \ref{eq:35}{]} preserves the effective mirror
symmetry $\hat{M}_{\hat{n}=\frac{\sqrt{2}}{2}\hat{x}+\frac{\sqrt{2}}{2}\hat{y}}=-\mathrm{i}\tau_{z}s_{z}\sigma_{z}$
with $\hat{M}_{\hat{n}}\mathcal{H}_{\mathrm{total}}^{\mathrm{SC}}(k_{x},k_{y})\hat{M}_{\hat{n}}^{-1}=\mathcal{H}_{\mathrm{total}}^{\mathrm{SC}}(k_{y},-k_{x})$
(where $\mathcal{H}_{\mathrm{total}}^{\mathrm{SC}}=\mathcal{H}_{\mathrm{BdG}}+\mathcal{H}_{\mathrm{SC}}$).
In fact, edge geometric phase mechanism is more suitable than crystalline symmetry
in the present situation for identifying the SC mass domain walls, since considering
the boundary pseudospin textures is more convenient than lattice symmetry analysis
in such complicated case. Here we can find the SC mass domain walls quickly and accurately
by our edge theory.

\end{document}